\documentclass[11pt,a4paper,useAMS,iop]{emulateapj}

\usepackage{tikz}
\usetikzlibrary{arrows,shapes,backgrounds}

\usepackage{xcolor}
\usepackage{threeparttable}
\usepackage{multirow}
\usepackage[colorlinks=true,linkcolor=blue,anchorcolor=blue,citecolor=blue,urlcolor=blue,draft=false]{hyperref}
\usepackage{amsmath}
\usepackage{graphicx}
\usepackage{morefloats}
\usepackage{natbib}
\usepackage[percent]{overpic}
\usepackage{enumerate}

\slugcomment{Submitted to ApJ. Draft version dated \today.}

\shorttitle{The Pre-Merging HST Frontier Fields Cluster MACS~J0416.1-2403}
\shortauthors{Ogrean et al.}

\newcommand{\chandra}{\emph{Chandra}}
\newcommand{\rosat}{\emph{ROSAT}}

\begin{document}


\title{Frontier Fields Clusters: Chandra and JVLA View of the\\Pre-Merging Cluster MACS J0416.1-2403}


\author{
G.~A.~Ogrean\altaffilmark{1,18}, R.~J.~van Weeren\altaffilmark{1,19}, C.~Jones\altaffilmark{1}, T.~E.~Clarke\altaffilmark{2}, 
J.~Sayers\altaffilmark{3}, T.~Mroczkowski\altaffilmark{2,20}, P.~E.~J.~Nulsen\altaffilmark{1}, W.~Forman\altaffilmark{1}, 
S.~S.~Murray\altaffilmark{4,1}, M.~Pandey-Pommier\altaffilmark{5}, S.~Randall\altaffilmark{1}, E.~Churazov\altaffilmark{6,7}, 
A.~Bonafede\altaffilmark{8}, R.~Kraft\altaffilmark{1}, L.~David\altaffilmark{1}, F.~Andrade-Santos\altaffilmark{1}, 
J.~Merten\altaffilmark{9}, A.~Zitrin\altaffilmark{3,18}, K.~Umetsu\altaffilmark{10}, A.~Goulding\altaffilmark{11,1}, 
E.~Roediger\altaffilmark{8,1,21}, J.~Bagchi\altaffilmark{12}, E.~Bulbul\altaffilmark{1}, M.~Donahue\altaffilmark{13}, 
H.~Ebeling\altaffilmark{14}, M.~ Johnston-Hollitt\altaffilmark{15}, B.~Mason\altaffilmark{16}, P.~Rosati\altaffilmark{17}, A.~Vikhlinin\altaffilmark{1}
}

\affil{\altaffilmark{}}
\affil{\altaffilmark{1}Harvard-Smithsonian Center for Astrophysics, 60 Garden Street, Cambridge, MA 02138, USA; \href{mailto:gogrean@cfa.harvard.edu}{gogrean@cfa.harvard.edu}}
\affil{\altaffilmark{2}U.S. Naval Research Laboratory, 4555 Overlook Ave SW, Washington, DC 20375, USA;}
\affil{\altaffilmark{3}Cahill Center for Astronomy and Astrophysics, California Institute of Technology, MC 249-17, Pasadena, CA 91125, USA;}
\affil{\altaffilmark{4}Department of Physics and Astronomy, Johns Hopkins University, 3400 N. Charles Street, Baltimore, MD 21218, USA;}
\affil{\altaffilmark{5}Centre de Recherche Astrophysique de Lyon, Observatoire de Lyon, 9 av Charles Andr\'e, 69561 Saint Genis Laval Cedex, France;}
\affil{\altaffilmark{6}Max Planck Institute for Astrophysics, Karl-Schwarzschild-Str. 1, 85741, Garching, Germany;}
\affil{\altaffilmark{7}Space Research Institute, Profsoyuznaya 84/32, Moscow, 117997, Russia;}
\affil{\altaffilmark{8}Hamburger Sternwarte, Universit\"at Hamburg, Gojenbergsweg 112 21029 Hamburg, Germany;}
\affil{\altaffilmark{9}Department of Physics, University of Oxford, Keble Road, Oxford OX1 3RH, UK;}
\affil{\altaffilmark{10}Institute of Astronomy and Astrophysics, Academia Sinica, PO Box 23-141, Taipei 10617, Taiwan;}
\affil{\altaffilmark{11}Department of Astrophysical Sciences, Princeton University, Princeton, NJ 08544, USA;}
\affil{\altaffilmark{12}Inter University Centre for Astronomy and Astrophysics,(IUCAA), Pune University Campus, Post Bag 4, Pune 411007, India;}
\affil{\altaffilmark{13}Department of Physics and Astronomy, Michigan State University, East Lansing, MI 48824, USA;}
\affil{\altaffilmark{14}Institute for Astronomy, University of Hawaii, 2680 Woodlawn Drive, Honolulu, HI 96822, USA;}
\affil{\altaffilmark{15}School of Chemical \& Physical Sciences, Victoria University of Wellington, PO Box 600, Wellington 6014, New Zealand}
\affil{\altaffilmark{16}National Radio Astronomy Observatory, 520 Edgemont Road, Charlottesville, VA 22903, USA;}
\affil{\altaffilmark{17}Department of Physics and Earth Science, University of Ferrara, Via G. Saragat, 1-44122 Ferrara, Italy;}


\altaffiltext{18}{Hubble Fellow}
\altaffiltext{19}{Einstein Fellow}
\altaffiltext{20}{National Research Council Fellow}
\altaffiltext{21}{Visiting Scientist}
\altaffiltext{}{}


\begin{abstract}
 Merging galaxy clusters leave long-lasting signatures on the baryonic and non-baryonic cluster constituents, including shock fronts, cold fronts, X-ray substructure, radio halos, and offsets between the dark matter and the gas components. Using observations from \chandra, the Jansky Very Large Array, the Giant Metrewave Radio Telescope, and the Hubble Space Telescope, we present a multiwavelength analysis of the merging Frontier Fields cluster MACS~J0416.1-2403 ($z=0.396$), which consists of a NE and a SW subclusters whose cores are separated on the sky by $\sim 250$~kpc. We find that the NE subcluster has a compact core and hosts an X-ray cavity, yet it is not a cool core. Approximately $450$~kpc south--south west of the SW subcluster, we detect a density discontinuity that corresponds to a compression factor of $\sim 1.5$. The discontinuity was most likely caused by the interaction of the SW subcluster with a less massive structure detected in the lensing maps SW of the subcluster's center. For both the NE and the SW subclusters, the dark matter and the gas components are well-aligned, suggesting that MACS~J0416.1-2403 is a pre-merging system. The cluster also hosts a radio halo, which is unusual for a pre-merging system. The halo has a $1.4$~GHz power of $(1.06\pm 0.09)\times 10^{24}$~W~Hz$^{-1}$, which is somewhat lower than expected based on the X-ray luminosity of the cluster. We suggest that we are either witnessing the birth of a radio halo, or have discovered a rare ultra-steep spectrum halo.

\end{abstract}


\keywords{Galaxies: clusters: individual: MACS J0416.1-2403 --- Galaxies: clusters: intracluster medium --- X-rays: galaxies: clusters}



\section{Introduction}

Galaxy clusters grow by merging with other clusters and by accreting smaller mass structures from the intergalactic medium. Signs of these interactions are imprinted in the ICM and detected in X-ray observations as cold fronts, shock fronts, turbulence, and ram pressure-stripped gas \citep[e.g.,][]{Markevitch2007,Randall2008a,Zhuravleva2015}. Other footprints of cluster interactions can be seen in the radio band as halos and relics \citep[e.g.,][]{Feretti2012}. If the merger is not in the plane of the sky, mergers can be detected in optical observations based on multiple peaks in the radial velocity distribution and in the spatial galaxy distribution. Furthermore, comparison of X-ray and optical/lensing data also reveals signatures of merging events, most notably as offsets between the dark matter (DM) and the gas components of the colliding clusters \citep[e.g.,][]{Markevitch2004,Clowe2006,Randall2008b,Merten2011,Dawson2012}.

Radio halos, which have been detected in some mergers, are diffuse synchrotron-emitting sources with low surface brightness, 
steep spectral indices ($\alpha<-1$, $\mathcal{F}_{\nu} \propto \nu^{\alpha}$), 
and typical sizes of $\sim 1$~Mpc. Because of their steep radio spectra, low-frequency radio 
observations play an important role in characterizing their properties.
Two main models have been proposed to explain the origin of such halos:
\begin{itemize}
	\item \emph{Reacceleration by turbulence}: Large-scale turbulence generated during the merger event supplies the energy required to reaccelerate fossil cosmic rays (CRs) back to relativistic energies, at which time they become synchrotron-bright \citep[e.g.,][]{Brunetti2001,Petrosian2001}.
	\item \emph{Acceleration by hadronic collisions}: Inelastic collisions between CR protons trapped in the gravitational potential of a cluster (e.g. originating from supernova explosions, active galactic nuclei [AGN] outbursts, previous merger events, etc.) and thermal ICM protons give rise to a secondary population of CR electrons, which consequently are visible at radio frequencies \citep[e.g.,][]{Blasi1999,Dolag2000}.
\end{itemize} 
However, hybrid models have also been postulated, in which radio halos may be produced by \emph{turbulent reacceleration of secondary particles} resulting from hadronic proton-proton collisions \citep{Brunetti2005,Brunetti2011}. The radio halos predicted by hybrid models are expected to be found in more relaxed clusters and to be underluminous for the masses of the hosting clusters \citep[see also][for a review]{Brunetti2014}.

Hadronic radio halo models predict that magnetic fields should either be different in clusters with and without halos, or that gamma-ray emission should be detected in clusters hosting halos \citep[e.g.,][]{Jeltema2011}. However, there is no evidence that magnetic fields are stronger in clusters with radio halos \citep[e.g.,][]{Bonafede2010}, and there has been no conclusive gamma-ray detection with the Fermi telescope \citep{Ackermann2010}. Therefore, current observational evidence disfavors hadronic models.

A textbook example of a merging cluster with a bright radio halo is the famous Bullet cluster, 1E~0657-56 \citep{Elvis1992,Liang2000,Markevitch2002,Shimwell2014}. \chandra\ has revealed that this system has a bullet-like gas cloud moving through the disturbed ICM of the main cluster \citep{Markevitch2002}. The gas cloud is the surviving core of a subcluster whose outer layers were ram pressure-stripped in the collision. Immediately ahead of the core there is a density discontinuity associated with a cold front, while further out in the same direction there is a density discontinuity associated with a shock front \citep{Markevitch2002,Owers2009}. Recently, another shock front has been found on the opposite side of the Bullet cluster, behind the bullet-like gas cloud \citep{Shimwell2015}. \chandra\ observations of the Bullet cluster were followed up with optical/lensing data acquired, most notably, with the \emph{Very Large Telescope} (VLT) and the \emph{Hubble Space Telescope} (HST). \citet{Chung2010} have shown that the redshift distribution of the Bullet cluster is bi-modal, with two redshift peaks at $z\sim 0.21$ and $z\sim 0.35$, which imply a velocity difference $\sim 3000$ km~s$^{-1}$ between the two merging subclusters. The gravitational lensing analysis has shown that the Bullet cluster is a dissociative merger, in which the essentially collisionless DM \citep[$\sigma_{\rm DM} < 0.7$~g~cm$^{-2}$,][]{Randall2008b} has decoupled from the collisional ICM \citep{Markevitch2004,Randall2008b}. \citet{Lage2014} have combined the X-ray and lensing data with numerical simulations to constrain the merger scenario of the Bullet cluster.

Here, we present results from \chandra, \emph{Jansky Very Large Array} (JVLA), and Giant Metrewave Radio Telescope (GMRT) observations of another merging galaxy cluster: the HST Frontier Fields\footnote{http://www.stsci.edu/hst/campaigns/frontier-fields/} cluster MACS~J0416.1-2403 \citep[z=0.396,][]{Ebeling2001,Mann2012}. Optical and lensing studies of this cluster have been presented recently by, e.g., \citet{Zitrin2013,Schirmer2014,Jauzac2014,Zitrin2015,Grillo2015}. Most recently, \citet{Jauzac2015} mapped the DM distribution of this merging system using strong- and weak-lensing data. Their analysis revealed two mass concentrations associated with the two main subclusters involved in the merger, plus two smaller X-ray-dark mass structures NE and SW of the cluster center. \citet{Jauzac2015} combined their lensing data with archival \chandra\ observations to study the offsets between the DM and the gas components. They report good DM-gas alignment for the NE subcluster, but a significant offset for the SW one. Based on the lensing and X-ray results, \citet{Jauzac2015} proposed two possible scenarios for the merger event in MACS~J0416.1-2403 -- one pre-merging and one post-merging -- but were unable to distinguish between the two. 

The analysis presented herein combines significantly deeper \chandra\ observations with recently acquired JVLA and GMRT data and with the optical/lensing results reported by \citet{Jauzac2015} to improve our understanding of the merger event in MACS~J0416.1-2403. The \chandra\, JVLA, and GMRT observations and data reduction procedures are described in Section~\ref{sec:data-reduction}. In Section~\ref{sec:bkg} we discuss the background analysis of the \chandra\ data, and in Sections~\ref{sec:global}, \ref{sec:mappings}, \ref{sec:subclusters}, and~\ref{sec:substructure} we present the X-ray results. The radio results are presented in Section~\ref{sec:jvla-results}. In Section \ref{sec:decoupling} we use the deeper \chandra\ data to revise the DM-gas offsets reported by \citet{Jauzac2015}. In Section~\ref{sec:scenario} we discuss the implications of our findings for the merger scenario of MACS~J0416.1-2403. In Section~\ref{sec:underluminous} we examine the origin of the radio halo. A summary of our results is provided in Section~\ref{sec:summary}.

Throughout the paper we assume a $\Lambda$CDM cosmology with $H_{\rm 0} = 70$~km~s$^{-1}$~Mpc$^{-1}$, $\Omega_{\rm M} = 0.3$, and $\Omega_{\rm \Lambda} = 0.7$. At the redshift of the cluster, 1\arcmin\ corresponds to approximately 320~kpc. Unless stated otherwise, the errors are quoted at the $90\%$ confidence level.

\section{Observations and Data Reduction}
\label{sec:data-reduction}

\subsection{Chandra}

The HST Frontier Fields cluster MACS J0416.1-2403 was observed with \chandra\ for $324$~ks between June 2009 and December 2014. A summary of the observations is given in Table~\ref{tab:spclean}. 

The data were reduced using CIAO v4.7 with the calibration files in CALDB v4.6.5. The particle background level of the VFAINT observations was lowered by filtering out cosmic ray events associated with significant flux in the 16 border pixels of the $5\times5$ event islands. Soft proton flares were screened out from all observations using the \emph{deflare} routine, which is based on the \emph{lc\_clean} code written by M. Markevitch. The flare screening was done using an off-cluster region at the edges of the field of view (FOV), which was selected to include only pixels located at distances greater than 2.5 Mpc from the cluster center and to exclude any point sources detectable by eye. The clean exposure times following this cleaning are listed in Table~\ref{tab:spclean}.

\begin{table*}
	\caption{\chandra\ observations}
	\label{tab:spclean}
	\begin{center}
		\begin{tabular}{llcccc}
	     	     \hline
			\multirow{2}{*}{ObsID} &  \multirow{2}{*}{Observing mode} &  \multirow{2}{*}{CCDs on} & \multirow{2}{*}{Starting date} & \multirow{2}{*}{Total time} & \multirow{2}{*}{Clean time} \\
					&  &  &  & (ks) & (ks) \\
	     	     \hline
			10446	& VFAINT   & $0, 1, 2, 3, 6$   & 06-07-2009   & 15.8   & 15.8 \\ 
			16236	& VFAINT   & $0, 1, 2, 3, 6$   & 08-31-2014   & 39.9   & 38.6 \\
			16237	& FAINT      & $0, 1, 2, 3, 6$   & 11-20-2013   & 36.6   & 36.3 \\
			16304	& VFAINT   & $0, 1, 2, 3, 6$   & 06-10-2014   & 97.8   & 95.2 \\
			17313	& VFAINT   & $0, 1, 2, 3, 6$   & 11-28-2014   & 62.8   & 59.9 \\ 
			16523	& VFAINT	  & $0, 1, 2, 3, 6$   &  12-19-2014  & 71.1   & 69.1 \\
	     	     \hline
		\end{tabular}
	\end{center}
\end{table*}

\begin{table*}[]
\begin{center}
\caption{JVLA L-band Observations}
\begin{tabular}{lllll}
& BnA-array & CnB-array & DnC-array \\
\hline
Observation dates &   25 Jan, 2014 & 12 Jan, 2015   & Sep 19, 2014  \\
Frequency coverage (GHz)           & 1--2 & 1--2 & 1--2 \\
On source time   (hr)   & 0.6 &  3.2 &  1.5 \\
Correlations   & full stokes & full stokes & full stokes \\
Channel width (MHz) &1  & 1 & 1 \\
Visibility integration time (s) & 2 & 3 & 5 \\
\hline
\end{tabular}
\label{tab:jvlaobs}
\end{center}
\end{table*}

\begin{table*}[]
\begin{center}
	\caption{Radio imaging parameters}
	\begin{tabular}{llcc}
		\hline
		Instrument    &\multirow{2}{*}{Weighting}	&	Resolution		&	r.m.s. noise		\\
			      &					&    (arcsec $\times$ arcsec)	&	($\mu$Jy beam$^{-1}$)	\\
		\hline
		JVLA 1.25~GHz & Briggs -0.75 			&	$7.8 \times 5.5$	&	14		\\			
		JVLA 1.25~GHz & uniform + 20\arcsec\ Gaussian taper&	$18 \times 18$		&	20	\\
		GMRT 610~MHz & Briggs 0.5 			&	$7.6 \times 4.0$	&	54		\\			
		GMRT 610~MHz & uniform + 20\arcsec\ Gaussian taper&	$ 20 \times 20$		&	$0.36\times10^3$	\\
        VLITE 340~MHz & Briggs 0.0 			&	$45.7 \times 31.1$	&	$2.1\times10^3$		\\			
		\hline
	\end{tabular}
	\label{tab:jvlaimages}
\end{center}
\end{table*}

\begin{figure*}[b!]
\begin{tikzpicture}
	\node[anchor=south west,inner sep=0] (image) at (0,0.0) {\includegraphics[width=0.9\textwidth]{{{0.5-4.0_mosaic_flux}.pdf}}};
            \begin{scope}[x={(image.south east)},y={(image.north west)}]
            	\path[draw=gray,style=very thick,densely dashed] (0.475,0.615) -- (0.675,0.615) -- (0.675,0.375) -- (0.475,0.375) -- (0.475,0.615); 
            	\node[anchor=south east] (image) at (1.1,0.8) {\includegraphics[width=0.6\textwidth]{{{0.5-4.0_zoomin_flux}.pdf}}};
		\path[draw=gray,style=very thick,densely dashed] (0.475,0.615) -- (0.535,1.500);
		\path[draw=gray,style=very thick,densely dashed] (0.675,0.615) -- (1.086,0.853);
	\end{scope}
\end{tikzpicture}
	\caption{Logarithmic-scaled \chandra\ surface brightness map of the HST Frontier Fields Cluster MACS~J0416.1-2403, in the energy band $0.5-4$~keV. The image was binned by a factor of 4 (1 pixel $\approx$ 2\arcsec), exposure-corrected, vignetting-corrected, and smoothed with a 2D Gaussian kernel of width 1 pixel $\times 1$ pixel. The large image shows the mosaic of all the ObsIDs summarized in Table~\ref{tab:spclean}, while the top image zooms in on the brightest X-ray regions of the cluster, in a region of size $2.9\times 2.9$~Mpc$^2$. The dashed grey circle marks the boundary of the $R_{\rm 500}$ region of the cluster.\label{fig:xray-mosaic}}
\end{figure*}
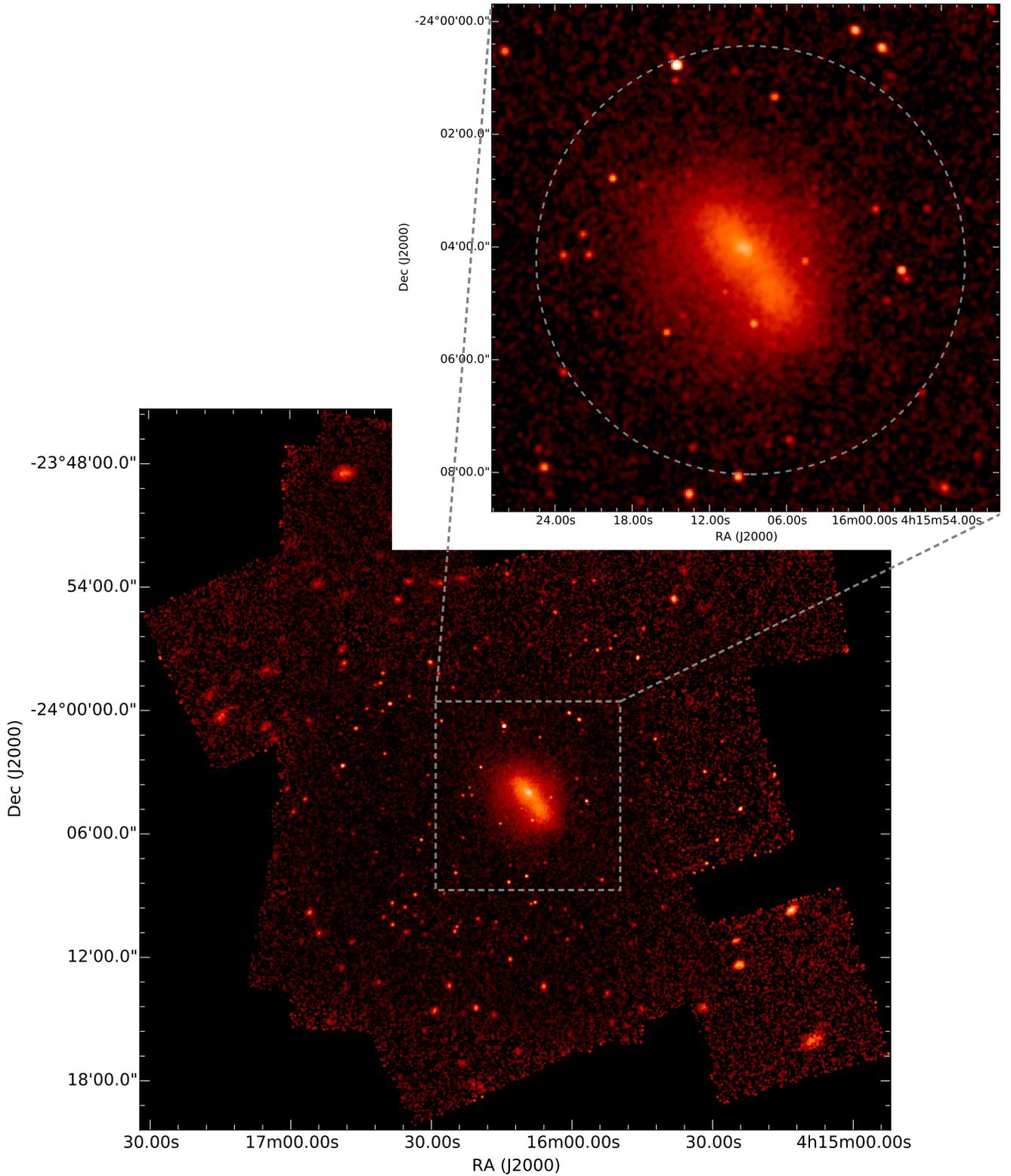


The inspection of an ObsID~10446 spectrum extracted from an off-cluster region showed a significant high-energy tail, which indicates that this observation is contaminated by flares that were not detected by the filtering routine. Given the relatively short exposure time of this observation, we decided to exclude ObsID~10446 from our analysis rather than attempt to model the flare component.

We reprojected the other five observations to a common reference frame, and created merged images in the energy bands $0.5-2$, $0.5-3$, $0.5-4$, $0.5-7$~keV, and $2-7$~keV. The exposure-corrected, vignetting-corrected $0.5-4$~keV mosaic image is shown in Figure \ref{fig:xray-mosaic}. To detect point sources, we also created merged exposure map-weighted PSF images (${\rm ECF}=90\%$) in each of the five energy bands. Point sources were detected in the individual bands with the CIAO task \emph{wavdetect}, using the merged maps, wavelet scales of 1, 2, 4, 8, 16, and 32 pixels, a sigma threshold of $5\times 10^{-7}$, and ellipses with $5\sigma$ axes. A few additional point sources that were missed by \emph{wavdetect} were selected by eye and excluded from the data. All point sources detected by \emph{wavdetect} were excluded from the data very conservatively, using the elliptical region that covered the largest area among the five elliptical regions identified for the different energy bands.

\subsection{JVLA}

JVLA observations of MACS~J0416.1-2403 were obtained in the L-band in the BnA, CnB, and DnC array configurations. 
The observations were recorded with the default wide-band L-band setup, giving 16 spectral windows 
each having 64 channels, covering the entire 1--2~GHz band. An overview of the observations is given in Table~\ref{tab:jvlaobs}.

The data were calibrated with the Common Astronomy Software Applications\footnote{http://casa.nrao.edu} 
(CASA) package version 4.2.1. As a first step we flagged data affected by radio frequency 
interference (RFI) using AOFlagger \citep{2010MNRAS.405..155O}, after correcting for the 
bandpass shape. Data affected by other problems and antenna shadowing were flagged as well. 
After flagging, we applied the pre-determined elevation dependent gain tables and antenna offset 
positions. The data were Hanning smoothed.

We determined initial gain solutions using 10 channels at the center of the spectral 
windows for the primary calibrators 3C147 and 3C138. We then re-calibrated the bandpass 
and obtained delay terms ({\tt gaintype=`K'}) using the unpolarized calibrator 3C147. 
A next step consisted of calibration of the cross-band delays ({\tt gaintype=`KCROSS'}) 
using the polarized calibrator 3C138. We then solved again for the gains on the primary 
calibrators but now using all channels. The channel dependent leakage corrections were 
found using 3C147 and the polarization angles were set using 3C138. The gains were again 
re-determined for all calibrator sources which included the phase calibrator J0416-1851. 
Finally, the flux density-scale was bootstrapped  from the primary calibrators to J0416-1851 
and the calibration solutions were applied to the target field.

To refine the calibration for the target field, we performed three rounds of 
phase-only self-calibration and two final rounds of amplitude and phase self-calibration.
W-projection  \citep{2008ISTSP...2..647C,2005ASPC..347...86C}  was employed during 
the imaging, taking the non-coplanar nature of the array into account.
The self-calibration was independently performed for the three different datasets 
from the different array configurations. The full bandwidth was imaged using 
MS-MFS clean \citep[{\tt nterms=2};][]{2011A&A...532A..71R}. \cite{briggs_phd} 
weighting with a {\tt robust} factor of -0.75 was used for self-calibration. 
Clean masks were used, which were made with the {PyBDSM\footnote{\url{http://dl.dropboxusercontent.com/u/1948170/html/index.html}}} 
source detection package.

After the self-calibration, the three datasets were combined. 
One final round of self-calibration was carried out on the combined dataset. 
The final images were corrected for the primary beam attenuation.

In addition, we imaged the dataset using an inner uv-range cut of 4.3~k$\lambda$ 
(corresponding to a scale of about 1\arcmin, or 320~kpc). This model was 
then subtracted from the visibility data to allow a search for diffuse emission 
in the cluster. We imaged the dataset with the emission from compact sources subtracted 
using a Gaussian uv-taper of 20\arcsec\ and employing multi-scale clean \citep{2008ISTSP...2..793C}. 
An overview of the resulting image properties, such as root-mean-square (rms) noise and resolution are given in Table~\ref{tab:jvlaimages}.

\subsection{GMRT}

GMRT 610~MHz data for MACS~J0416.1-2403 were 
collected on December 3, 2013 using 26 antennas. The on-source time was 5.0~hrs, 
and a total bandwidth of 33.3~MHz (RR correlations only), split into 512 channels, was recorded. 
NRAO's Astronomical Image Processing System (AIPS) was used to carry out the initial calibration of  
the visibility dataset. The primary calibrator 3C 147 was used to set the flux density scale 
and derive the bandpass solutions for all the antennas. The source J0409-179 was observed for 6-mins scans 
at intervals of $\sim25$~mins and used as the secondary phase and gain calibrator. 
About 20\% of the data, mostly from short baselines, were affected by RFI and subsequently flagged. 
Gain solutions were obtained for the calibrator sources and together with the bandpass solutions 
applied to the target field. In total, 480 of the 512 channels 
were used; the rest of the channels were discarded as they were too noisy due to the bandpass  roll-off.
The 480 channels were averaged down to 48 channels to reduce the size of the data.

The visibility data were then imported into CASA to refine the calibration via the process of self-calibration.
Three rounds of phase-only self-calibration and two final rounds of
amplitude and phase self-calibration were applied. The phase-only self-calibration was carried out on a 30~s timescale. 
The amplitude and phase self-calibration was carried out on a 2~min timescale, 
pre-applying the phase-only solutions before solving for the amplitude and phases on the longer 2~min timescale.
We employed W-projection during the imaging and used {\tt nterms=2}.

A map with Briggs ({\tt robust=0.5}) weighting of the field was produced, which resulted in a resolution of  
$7.6\arcsec \times 4.0\arcsec$ and an rms noise level of 54~$\mu$Jy~beam$^{-1}$ (Table~\ref{tab:jvlaimages}). 
Similarly to the JVLA L-band data reduction, we imaged the dataset using an inner uv-range cut 
of 4.3~k$\lambda$ to obtain a model of the compact sources in the field. After subtracing this model from the uv-data, 
we re-imaged with uniform weighting and a Gaussian uv-taper to obtain a beam size of 20\arcsec.

\subsection{VLITE}

A new commensal observing system called the VLA Low Band Ionospheric
and Transient Experiment (VLITE)\footnote{\url{http://vlite.nrao.edu/}} has been developed for the NRAO JVLA
(Clarke et al., in prep.) and was operational during the observations
of MACS~J0416.1-2403.   The VLITE correlator is a custom
designed DiFX software correlator \citep{2011PASP..123..275D}.  The
system processes 64~MHz of bandwidth centered on 352 MHz with two
second temporal resolution and 100~kHz spectral resolution.
VLITE operates during nearly all pointed
JVLA observations with primary science goals at frequencies above 1~GHz,
providing data simultaneously for 10 JVLA antennas using the low band
receiver system \citep{6051196}.

We processed the CnB configuration VLITE observation of MACS~J0416.1-2403 
from 12 January 2015 using a combination of the Obit software package 
\citep{2008PASP..120..439C} and AIPS \citep{1996ASPC..101...37V}. 
VLITE data at frequencies $\nu>$360~MHz were removed due to the presence of strong
RFI from a satellite downlink that is present during most operations.
The data at $\nu<$360~MHz were flagged using the AIPS program
RFLAG to remove the majority of the remaining RFI. The bandpass 
was flattened using observations of several calibrators taken near the
time of the observations (3C48, 3C138, 3C147, and 3C286). These
calibrator sources were taken in the same primary observing band as
MACS~J0416.1-2403. The delays were determined using these same
calibrators. The delay corrected data then were flux calibrated using
these same calibrator sources. No phase calibrator is required for low
band observations due to the large field of view (FWHM $\sim$ 2.3$^\circ$
 at 320~MHz) and typical presence of sufficiently strong sources
in the field that allow for self-calibration. 

The initial imaging of the target field revealed additional RFI that
was identified in the residual data set after all compact sources had
been subtracted from the $uv$ data. These baselines were excised from
the full data set and we further refined the calibration through four
rounds of phase-only self-calibration. Non-coplanar effects were taken
into account in the imaging steps using small facets to make a fly-eye
image of the full field out to the FWHM with additional facets placed
on bright sources out to 20$^\circ$ from the field center. Data were
imaged using small clean masks placed around all sources.
The final VLITE image of MACS~J0416.1-2403 has a resolution of 45.7\arcsec\
$\times$ 31.1\arcsec\ and an rms noise of 2.1~mJy~beam$^{-1}$. The image
was made using a Briggs robust parameter of 0 (see Table~\ref{tab:jvlaimages}).

\section{X-ray Foreground \& background modelling}
\label{sec:bkg}

We analyzed the \emph{Chandra} data using the Group F stowed background event files, which are appropriate for observations taken after September 21, 2009. The particle background was subtracted from the data, while the foreground and background sky components were modeled from off-cluster regions. For each of the cluster ObsIDs, we created corresponding stowed background event files, applied the VFAINT cleaning to those associated with cluster observations taken in this observing mode, and renormalized the stowed background observations such that the count rates in the energy band $10-12$~keV matched those of the corresponding observation in the same energy band. We modeled the sky components as the sum of unabsorbed thermal emission from the Local Hot Bubble \citep[LHB; APEC component with $T\approx 0.1$~keV;][]{Snowden1998}, absorbed thermal emission from the Galactic Halo \citep[GH; APEC component with $T\approx 0.2$~keV;][]{Henley2010}, and absorbed non-thermal emission from undetected point sources in the FOV \citep[power-law component with index $1.41\pm 0.06$;][]{DeLuca2004}. The redshifts of the foreground components were set to 0, while the abundances were set to solar values assuming the abundance table of \citet{angr1989}. The hydrogen column density was fixed to $2.89\times 10^{20}$~atoms~cm$^{-2}$ \citep{Kalberla2005}. The \chandra\ spectra were fit in the energy band $0.5-7$ keV using \textsc{Xspec} v12.8.2 -- the lower limit was chosen to avoid calibration uncertainties at low energies, and the upper limit to increase the signal-to-noise ratio (SNR). 

When fitting the \chandra\ spectra in the energy band $0.5-7$~keV, the $\sim 0.1$~keV LHB component cannot be constrained. Instead, we constrain this component using a ROentgen SATellite (\rosat) All-Sky Survey (RASS) spectrum\footnote{\url{https://heasarc.gsfc.nasa.gov/cgi-bin/Tools/xraybg/xraybg.pl}} corresponding to an annulus with radii 0.15 and 1.0 degrees around the cluster. The \rosat\ spectrum was fit in the energy band $0.1-2.4$~keV, with the normalization of the power-law background component fixed to $8.85\times 10^{-7}$ photons~keV$^{-1}$~cm$^{-2}$~s$^{-1}$~arcmin$^{-2}$ \citep{Moretti2003}. The temperatures and normalizations of the LHB and GH components were free in the fit. The results of the fit to the \rosat\ spectrum are presented in Table~\ref{tab:bkg-model}.

Based on the \rosat\ results, the parameters of the LHB components were fixed for the \chandra\ spectra. The five \chandra\ spectra were then fitted in parallel, assuming that they are described by the same foreground model. The power-law normalizations, on the other hand, were left to vary independently, in order to account for the varying exposure time across the merged observation that was used for point source detection. The energy sub-band $0.5-0.8$~keV was ignored for ObsID 16237 due to negative spectral residuals caused by a change in the spectral shape of the background component that is not filtered by the VFAINT cleaning, which occurred between 2009 (the date of the Group F background files) and 2013 (the date of the observation; A. Vikhlinin, priv. comm.). The subtraction of the instrumental background resulted in small line residuals near the spectral positions of the Al~K$\alpha$ ($E\approx 1.49$~keV), Si~K$\alpha$ ($E\approx 1.75$~keV), and Au~M$\alpha,\beta$ ($E\approx 2.1$~keV) fluorescent instrumental lines. Therefore, we excluded from the spectra very narrow bands ($\Delta E=0.10$~keV) surrounding these lines.

The best-fitting foreground and background parameters are summarized in Table~\ref{tab:bkg-model}. The spectra were grouped to have at least 1 count per bin, and the fit was done using the extended C-statistic\footnote{\url{https://heasarc.gsfc.nasa.gov/xanadu/xspec/manual/XSappendixStatistics.html}} \citep{Cash1979}. 


\begin{table*}
	\caption{Foreground and background spectral models}
	\label{tab:bkg-model}
	\begin{center}
	     \begin{threeparttable}
		\begin{tabular}{lcccc}
	     	     \hline
		     	       & \multicolumn{2}{c}{\chandra} & \multicolumn{2}{c}{\rosat} \\
		     \hline
		     	Component & $T$\tnote{a} & $\mathcal{N}$\tnote{b} & $T$\tnote{a} & $\mathcal{N}$\tnote{b} \\
		     \hline
		      	LHB & $0.10\pm 0.01$\tnote{c} & $9.32_{-0.83}^{+0.62}\times 10^{-7}$\tnote{c} & $0.10\pm 0.01$ & $9.32_{-0.83}^{+0.62}\times 10^{-7}$ \\
			GH & $0.25_{-0.08}^{+0.23}$ & $2.35_{-1.66}^{+4.20}\times 10^{-7}$ & $0.22_{-0.04}^{+0.05}$ & $6.74_{-2.06}^{+3.44}\times 10^{-7}$ \\
			power-law & \multirow{2}{*}{--} & \multirow{2}{*}{$4.20_{-0.73}^{+0.86}\times 10^{-7}$} & -- & -- \\
				\hspace{0.25cm} ObsID 16236 & & & & \\
			power-law & \multirow{2}{*}{--} & \multirow{2}{*}{$2.17_{-0.86}^{+0.76}\times 10^{-7}$} & -- & -- \\
				\hspace{0.25cm} ObsID 16237 & & & & \\
			power-law & \multirow{2}{*}{--} & \multirow{2}{*}{$2.45_{-0.59}^{+0.60}\times 10^{-7}$} & -- & -- \\
				\hspace{0.25cm} ObsID 16304 & & & & \\
			power-law & \multirow{2}{*}{--} & \multirow{2}{*}{$3.70_{-0.84}^{+0.86}\times 10^{-7}$} & -- & -- \\
				\hspace{0.25cm} ObsID 17313 & & & & \\
			power-law & \multirow{2}{*}{--} & \multirow{2}{*}{$4.33_{-0.69}^{+0.71}\times 10^{-7}$} & -- & -- \\
				\hspace{0.25cm} ObsID 16523 & & & & \\
	     	     \hline
		\end{tabular}
		\begin{tablenotes}
			\item[a] Temperature, in units of keV.
			\item[b] Normalization, in units of cm$^{-5}$~arcmin$^{-2}$ for the thermal components, and in units of photons~keV$^{-1}$~cm$^{-2}$~s$^{-1}$~arcmin$^{-2}$ for the non-thermal components.
			\item[c] Fixed parameter.
		\end{tablenotes}
	     \end{threeparttable}
	\end{center}
\end{table*}

\section{Global X-Ray Properties}
\label{sec:global}

MACS~J0416.1-2403 has been previously identified as a merging galaxy cluster at $z=0.396$ \citep{Mann2012}. The brightness of the NE subcluster core is significantly more peaked than that of the SW subcluster, which instead has a rather flat brightness distribution (Figure~\ref{fig:xray-mosaic}). 

We determined the average cluster temperature by extracting spectra from a circular region of radius $1.2$~Mpc \citep[approximately $R_{\rm 500}$,][]{Sayers2013} centered at $\alpha=4^{\rm h}\, 16^{\rm m}\, 08.8^{\rm s}$ and $\delta=-24^{\circ}\, 04\arcmin \, 14.0\arcsec$ (J2000). The spectra extracted from the five \chandra\ ObsIDs were fit in parallel, assuming the cluster emission is described by a single-temperature thermal component\footnote{We tried adding an additional thermal component to fit the cluster emission, but the parameters of the second component could not be constrained.}. For this global fit, the sky background parameters were fixed to the values listed in Table~\ref{tab:bkg-model}, and the redshift was fixed to $0.396$. We found $T=10.06_{-0.49}^{+0.50}$~keV and $L_{\rm X, \,\, 0.1-2.4\,\,keV} = (9.14\pm 0.10)\times 10^{44}$~erg~s$^{-1}$. The average cluster parameters are shown in Table~\ref{tab:global}.

\begin{table}
	\centering
	\caption{Cluster properties in $R_{\rm 500}$ 	\label{tab:global}}
	\begin{tabular}{lc}
		\hline
			T  (keV) &   $10.06_{-0.49}^{+0.50}$ \\
			Z (${\rm Z}_\sun$) & $0.24_{-0.04}^{+0.05}$ \\
			$L_{\rm X,\, 0.1-2.4\: {\rm keV}}\; {\rm (erg\,\,s^{-1})}$ & $\left(9.14\pm 0.10\right)\times 10^{44}$\\
		\hline
	\end{tabular}
\end{table}

\section{Mapping the Cluster's X-Ray Properties}
\label{sec:mappings}

\subsection{Mapping of the ICM Temperature}

We mapped the cluster properties using \textsc{contbin} \citep{Sanders2006}.\footnote{We also created temperature maps using the codes described by \citet{Churazov2003} and \citet{Randall2008a}, and obtained consistent results at the $90\%$ confidence level.} The merged \chandra\ image was adaptively binned in regions of $\sim 3600$ source plus sky background counts in the energy band $0.5-7$~keV. The bins follow the surface brightness contours of the cluster, thus minimizing possible gas mixing in bins located near density discontinuities. Instrumental background and total spectra were extracted from regions corresponding to each bin, and grouped to have at least one count per bin. The net spectra were modelled as the sum of sky background components plus a single temperature absorbed thermal model (\emph{phabs}$\times$\emph{APEC}) with free temperature and normalization. The metallicity was fixed to $Z=0.24Z_\sun$ (see Section~\ref{sec:global}). The sky background emission was fixed to the model summarized in Table~\ref{tab:bkg-model}. The spectra extracted from the five \chandra\ datasets were fitted in parallel, with the temperatures and normalizations linked between the different ObsIDs. The fits were done using the extended C-statistic. The resulting temperature map is shown in Figure~\ref{fig:global_map_fixedZ}.

The temperature is high throughout the cluster, which causes the uncertainties on the measurements to be high. The temperature map does not show significant structure. Within the uncertainties, the temperatures out to $\sim 400$~kpc from the cluster center are consistent with $\sim 10$~keV.

\subsection{Mapping of the ICM Pressure and Entropy}

The electron number density can be derived either from the model normalization of the fitted spectrum, or from the surface brightness in the regions of interest. Extracting the number density from either of these quantities requires assumptions about the cluster geometry. Here, we estimated the electron number density from the surface brightness; we note that our results are unchanged when deriving the density from the spectral normalization. 

The surface brightness is essentially proportional to the emission measure,
\begin{eqnarray}
	\zeta  \propto \int n_{\rm e}^2 {\rm d}l
\end{eqnarray}
where the integration is done along the line of sight and $n_{\rm e}$ is the electron number density. More accurately, the surface brightness is also temperature-dependent. This dependence introduces an uncertainty of $\lesssim 10\%$ in the pressure and entropy maps, which is less than the uncertainties on the temperatures and does not affect our results.

\begin{figure*}
	\centering
	\includegraphics[width=0.32\textwidth]{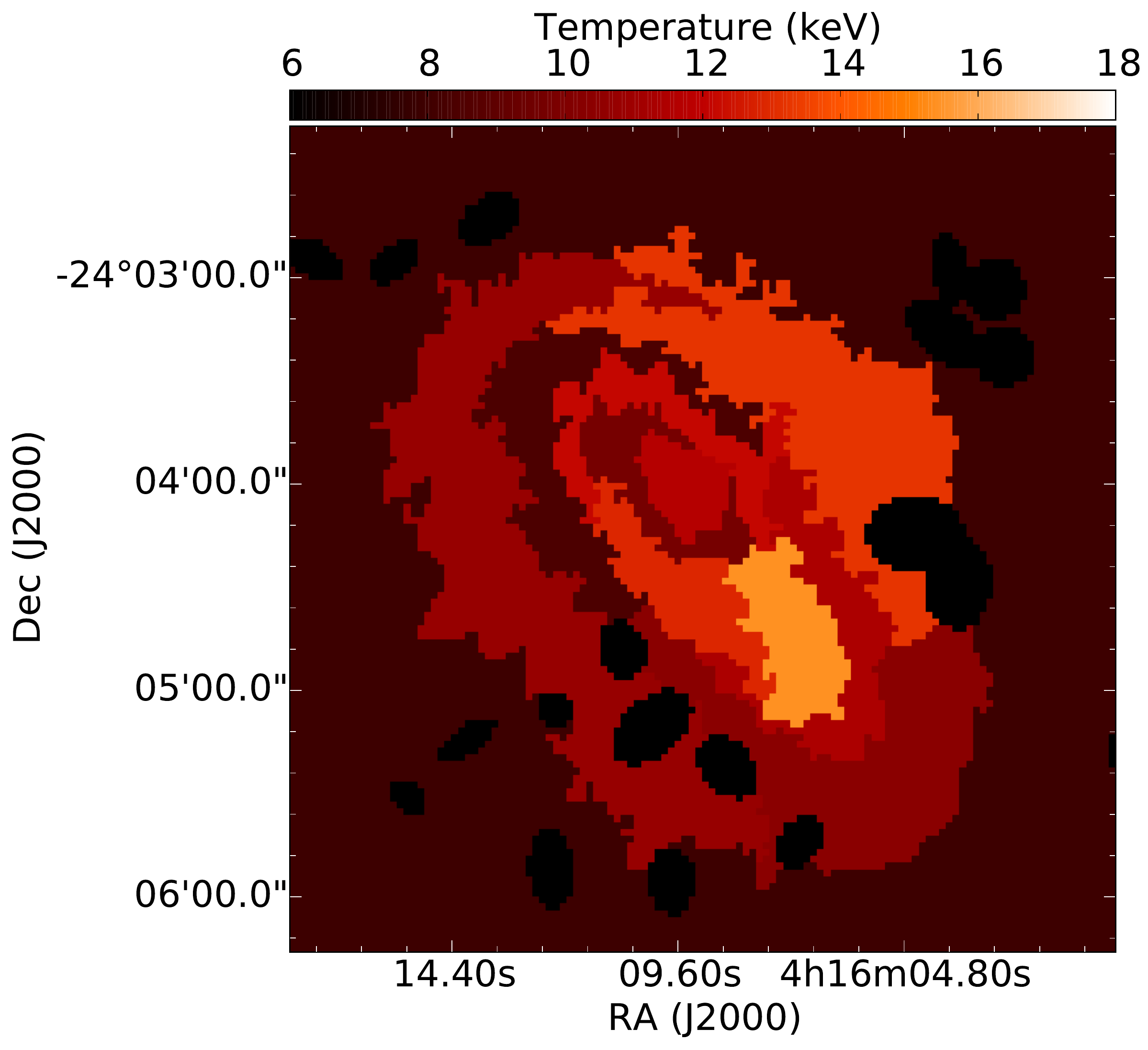}
	\includegraphics[width=0.32\textwidth]{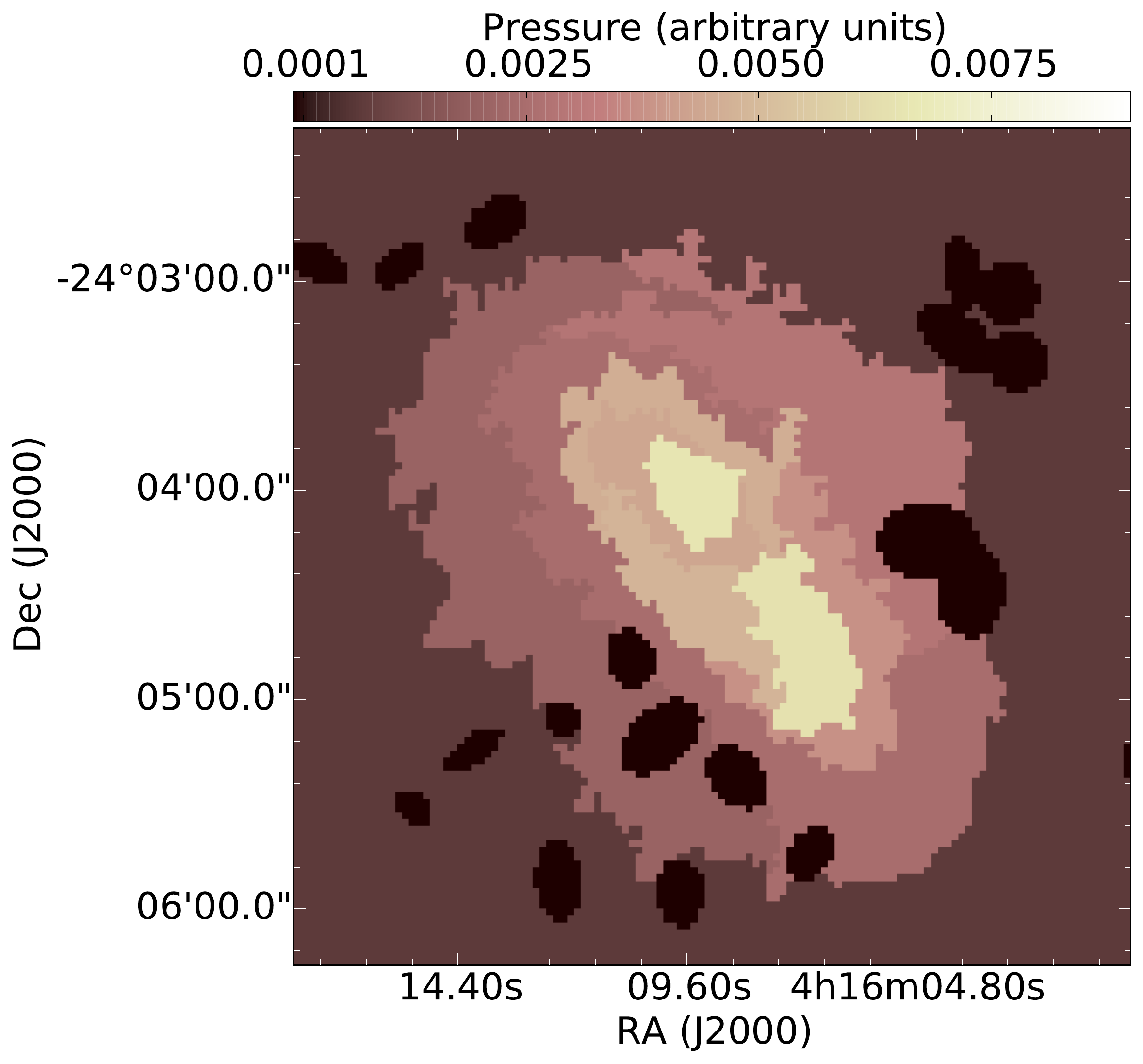}
	\includegraphics[width=0.32\textwidth]{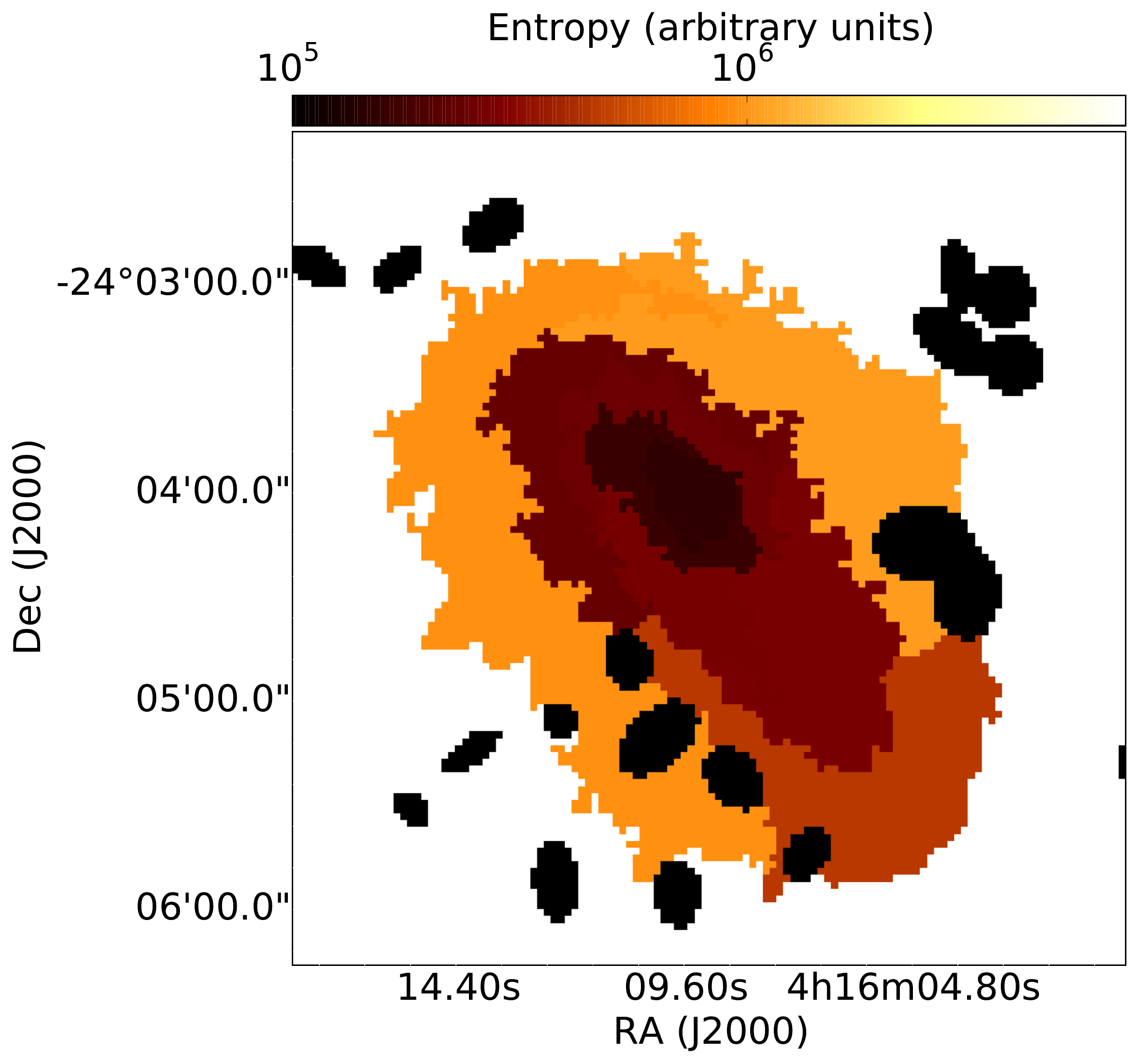}
	\caption{Temperature, pseudo-pressure, and pseudo-entropy maps of MACS~J0416.1-2403. In all maps, black elliptical regions mark the regions from which point sources have been removed.}
	\label{fig:global_map_fixedZ}
\end{figure*}

We approximated the density as $\zeta^{1/2}$, and defined the pseudo-pressure and pseudo-entropy as:
\begin{eqnarray}
	P & = & T \,\zeta_{0.5-2 \:{\rm keV}}^{1/2}\,\,, \\
	K & = & T \,\zeta_{0.5-2 \:{\rm keV}}^{-1/3}\,\,.
\end{eqnarray}
The pseudo-pressure and pseudo-entropy maps are included in Figure~\ref{fig:global_map_fixedZ}. 

Strong jumps in pressure and entropy would indicate the presence of strong density discontinuities in the ICM. However, neither the pressure map, nor the entropy map of MACS~J0416.1-2403 shows evidence of such strong jumps between the regions. We note that while a pressure and entropy discontinuity can be seen between regions $\#3$ and $\#0$, the size of region $\#0$ is too large for this result to truly indicate that there is a density discontinuity at the boundary of the two regions.

\begin{figure*}
	\centering
	\includegraphics[width=\columnwidth]{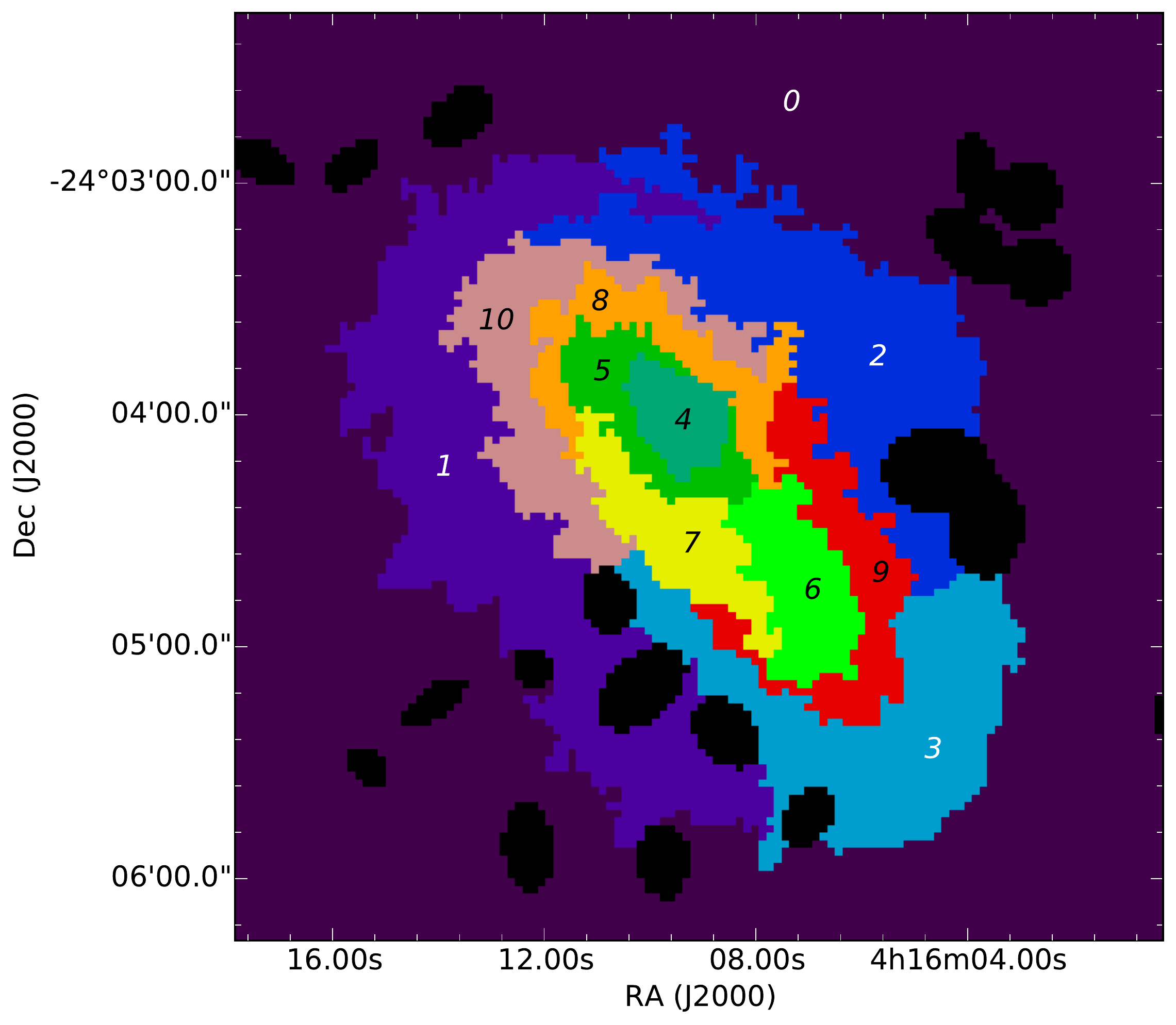}
	\includegraphics[width=\columnwidth]{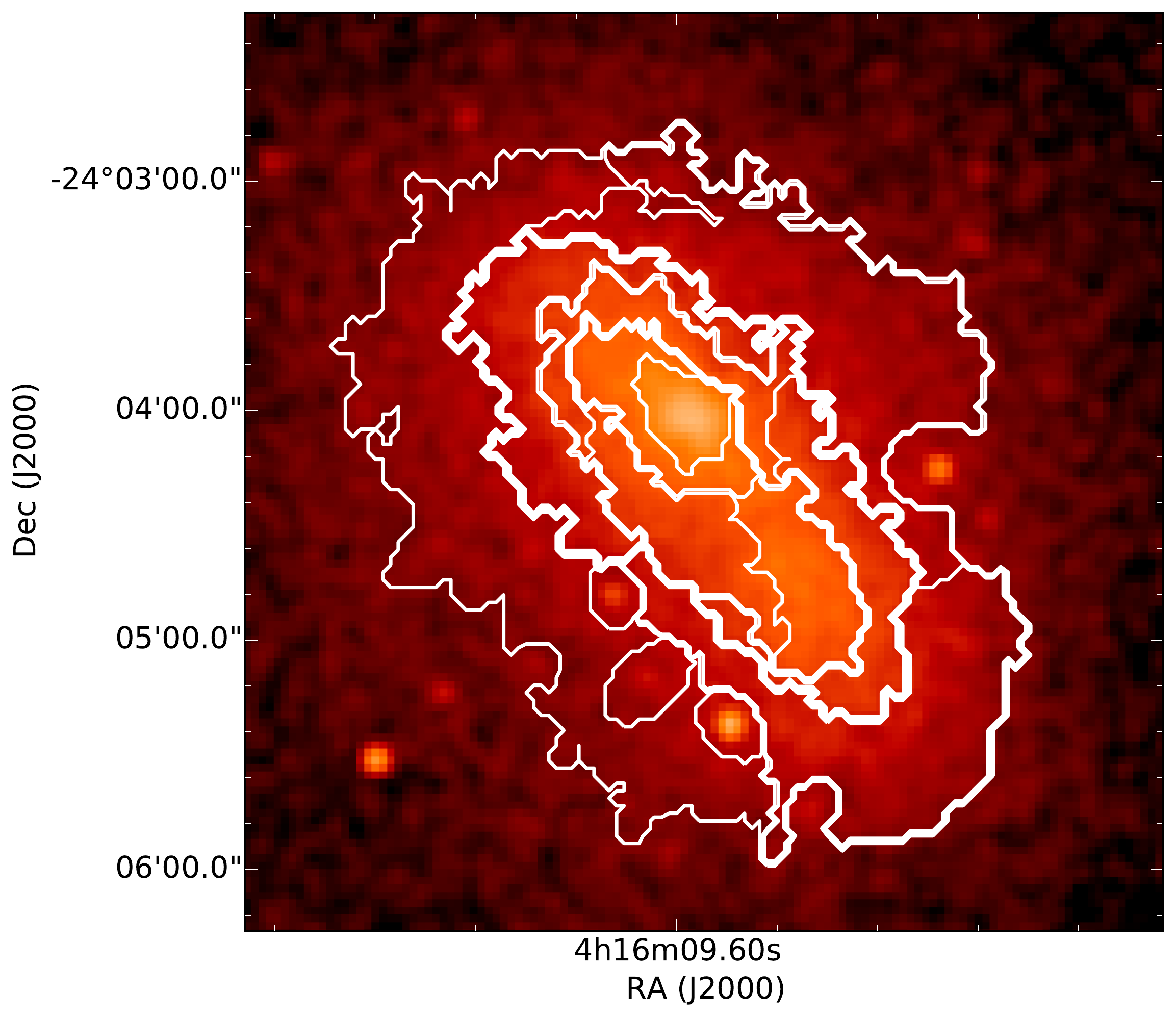}
	\caption{\emph{Left:} The map shows the association of regions in Figure~\ref{fig:global_map_fixedZ} with the fit values in Table~\ref{tab:map-fit-results}. Numbers are region numbers. Black ellipses mark the regions where point sources have been removed. \emph{Right:} Same surface brightness map as in Figure~\ref{fig:xray-mosaic}, with overlaid regions used for mapping the physical properties of the ICM.}
	\label{fig:binmap}
\end{figure*}

In Table~\ref{tab:map-fit-results} we list the best-fitting temperatures, spectral normalizations, and $0.5-2$~keV count rates for the regions in Figure~\ref{fig:global_map_fixedZ}. The bin numbers necessary to relate Table~\ref{tab:map-fit-results} and Figure~\ref{fig:global_map_fixedZ} are shown in Figure~\ref{fig:binmap}. An interactive version of the temperature map, which combines Figures~\ref{fig:global_map_fixedZ}, \ref{fig:binmap} and Table~\ref{tab:map-fit-results} is available at \url{http://hea-www.cfa.harvard.edu/~gogrean/interactive/MACSJ0416_Tmap.html}.

\begin{table*}
	\caption{Best-fitting temperatures, normalizations, and $0.5-2$~keV count rates for the regions in Figure~\ref{fig:binmap}.}
	\label{tab:map-fit-results}
	\begin{center}
	     \begin{threeparttable}
		\begin{tabular}{lccc}
	     	     \hline
		     	Bin number & $T$\tnote{a} & $\mathcal{N}$\tnote{b} & $S_{\rm X}$\tnote{c} \\
		     \hline
			0 & $   7.95_{-0.92}^{+1.32}$ & $(2.24 \pm 0.06) \times 10^{-5}$ & $(7.39\pm 0.07) \times 10^{-6}$ \\
			1 & $10.74_{-1.37}^{+2.10}$ & $(1.70\pm 0.05) \times 10^{-4}$ & $(3.38\pm 0.06) \times 10^{-5} $ \\
			2 & $13.25_{-2.32}^{+3.07}$ & $(2.31\pm 0.07) \times 10^{-4}$ & $(4.34\pm 0.08) \times 10^{-5}$ \\
			3 & $10.32_{-1.35}^{+1.91}$ & $(2.77\pm 0.08) \times 10^{-4}$ & $(5.30\pm 0.09) \times 10^{-5}$ \\
			4 & $11.69_{-1.46}^{+2.10}$ & $(1.63\pm 0.05) \times 10^{-3}$ & $(2.96\pm 0.06) \times 10^{-4}$ \\
			5 & $9.73_{-1.24}^{+1.52}$ & $(9.80\pm 0.29) \times 10^{-4}$ & $(1.84\pm 0.04) \times 10^{-4}$ \\
			6 & $15.39_{-2.87}^{+3.42}$ & $(9.03\pm 0.23) \times 10^{-4}$ & $(1.62\pm 0.03) \times 10^{-4}$ \\
			7 & $12.90_{-2.15}^{+3.64}$ & $(7.13\pm 0.22) \times 10^{-4}$ & $(1.27\pm 0.03) \times 10^{-4}$ \\
			8 & $12.18_{-1.90}^{+2.41}$ & $7.10_{-0.21}^{+0.22}\times 10^{-4}$ & $(1.31\pm 0.03) \times 10^{-4}$ \\
			9 & $11.43_{-1.65}^{+2.29}$ & $(5.51\pm 0.17)\times 10^{-4}$ & $(1.02\pm 0.02) \times 10^{-4}$ \\
			10 & $8.41_{-0.90}^{+1.37}$ & $(4.12\pm 0.13) \times 10^{-4}$ & $(8.02\pm 0.15)\times 10^{-5}$ \\
		     \hline
		\end{tabular}
		\begin{tablenotes}
			\item[a] Temperature, in units of keV.
			\item[b] Spectral normalization, in units of ${\rm cm^{-5}\;  arcmin^{-2}}$.
			\item[c] $0.5-2$~keV count rate, in units of ${\rm photons\; cm^{-2}\; s^{-1}\; arcmin^{-2}}$.
		\end{tablenotes}
	     \end{threeparttable}
	\end{center}
\end{table*}

\section{Properties of the Individual Subclusters}
\label{sec:subclusters}

The merging history of the individual subclusters is closely related to their degree of disturbance. In the following two sections, we perform the imaging analysis of the NE and SW subclusters in order to characterize their merging states. For both subclusters, we created surface brightness profiles in sectors centered on the respective X-ray peaks, and attempted to model the profiles with isothermal $\beta$-models \citep{Cavaliere1976,Cavaliere1978}:
\begin{eqnarray}
	S(r) = S_0 \left[1+\left(\frac{r}{r_c}\right)^2 \right]^{-3\beta+0.5} 
	\label{eq:beta}
\end{eqnarray}
where $S_0$ is the central surface brightness, $r_c$ is the core radius, and $r$ is the radius from the cluster centre.

Simple $\beta$-models provide good descriptions of the X-ray profiles of galaxy clusters that do not have strong ICM temperature gradients \citep[e.g., ][]{Ettori2000a,Ettori2000b}. Based on Figure~\ref{fig:global_map_fixedZ}, there are indeed no strong temperature gradients in the NE and the SW subclusters of MACS~J0416.1-2403.

Before fitting, all the profiles were binned to a minimum of 1 count/bin. The fits used Cash statistics. Fitting was done using a modified version of the PROFFIT package\footnote{Available upon request.} \citep{proffit}. All the surface brightness profiles presented in the following subsections are in the energy band $0.5-4$~keV. The profiles were instrumental background-subtracted, and exposure- and vignetting-corrected.

\subsection{NE Subcluster}

\begin{figure}
	\includegraphics[width=0.9\columnwidth]{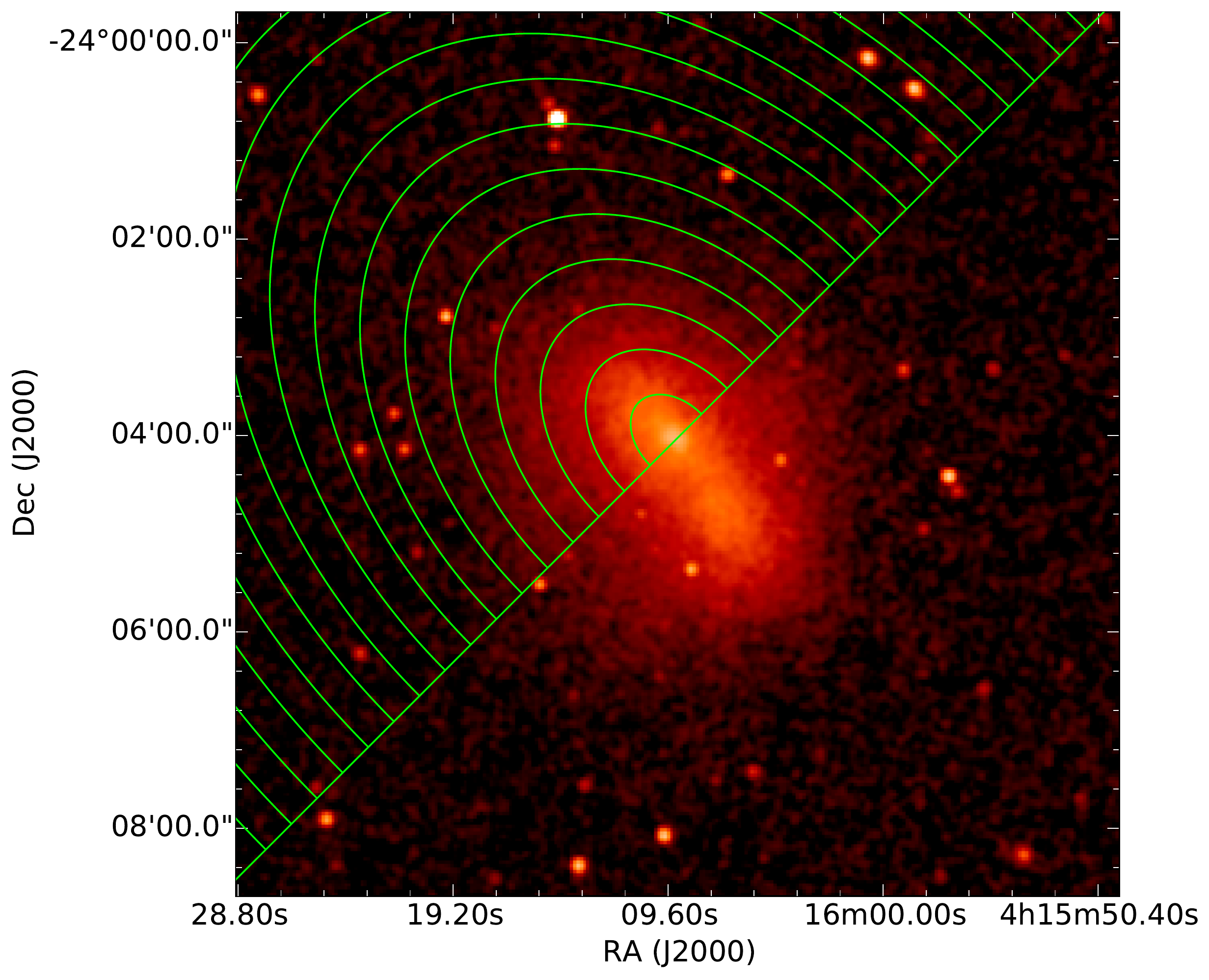}
	\includegraphics[width=\columnwidth]{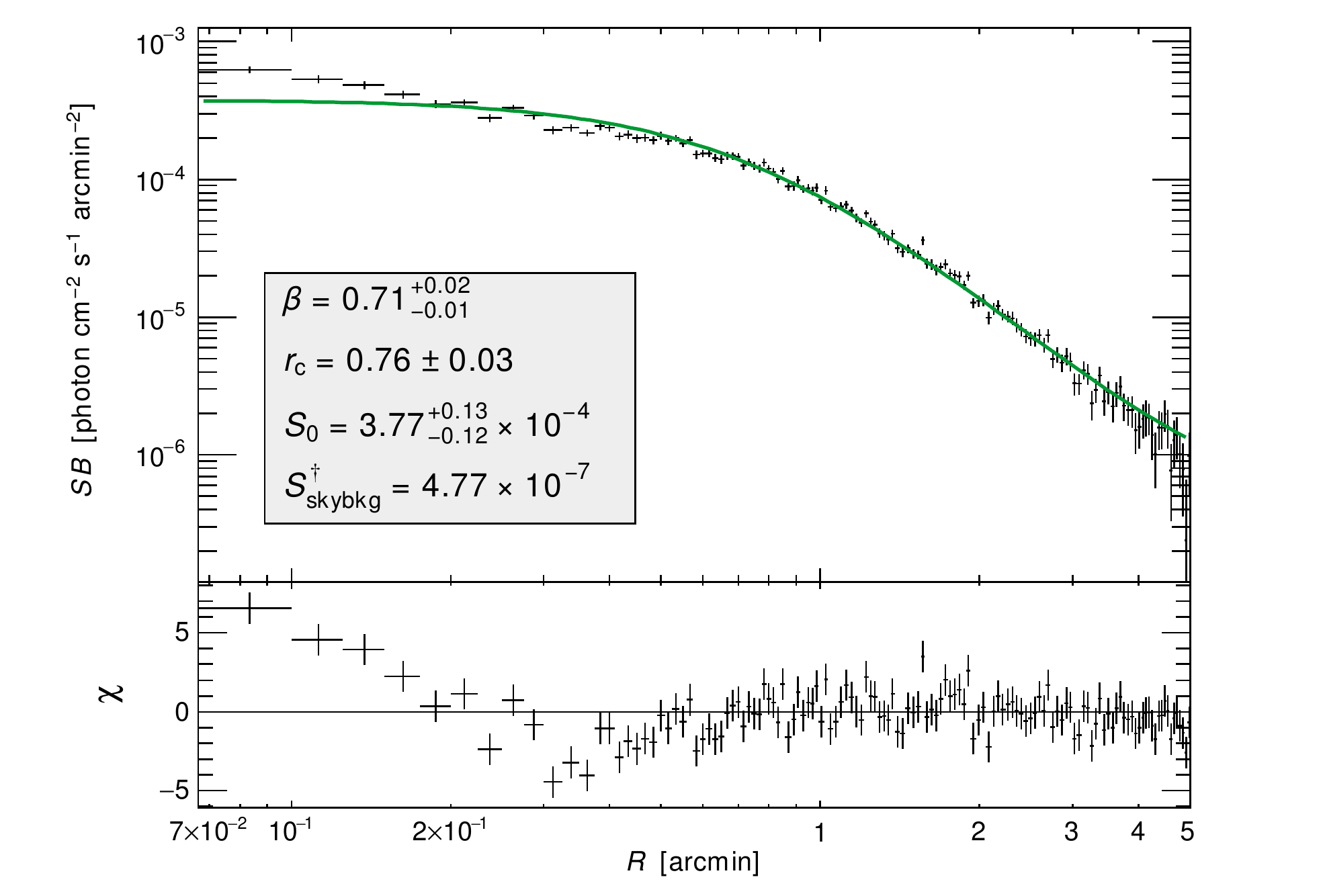}
	\includegraphics[width=\columnwidth]{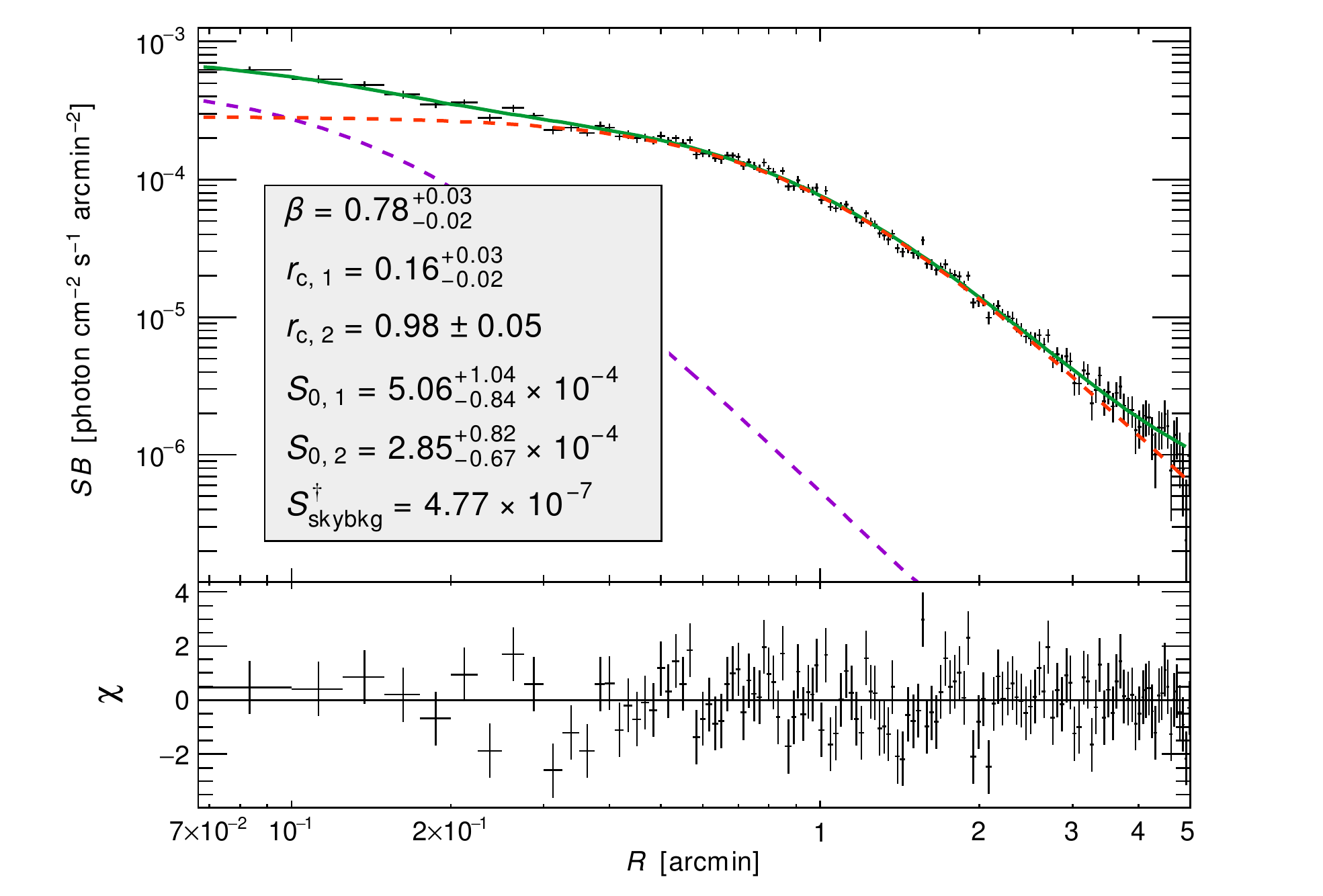}
	\caption{\emph{Top:} Sector used to model the surface brightness of the NE subcluster. Annuli are drawn only to guide the eye and do not reflect the actual bin size used for the surface brightness profiles. \emph{Middle:} $\beta$-model fit to the surface brightness profile of the NE subcluster. For clarity, the profile shown in this plot was binned to a uniform signal-to-noise ratio of 5. \emph{Bottom:} Double $\beta$-model fit to the surface brightness profile of the NE subcluster. The individual $\beta$-models are shown with dashed lines. For clarity, the profile shown in this plot was binned to a have 200 counts per bin. The best-fitting model parameters are listed in the middle and bottom plots; radii units are arcmin, and surface brightness units are photon~cm$^{-2}$~s$^{-1}$~arcmin$^{-2}$. Fixed parameters are shown with a superscripted $\dagger$. The bottom panel shows the residuals of the fit. \label{fig:sectors-N}}
\end{figure}

The sector used to create the surface brightness profile of the NE subcluster and the fits to this profile are shown in Figure~\ref{fig:sectors-N}. The sky background surface brightness level was determined by fitting a constant to the outer bins (radii $5.0\arcmin-10.7\arcmin$) of the profile. The sky background level was kept fixed in the following fits. The very central part of the profile is significantly underestimated by the best-fitting $\beta$-model. Instead, the profile is described better by a double $\beta$-model:
\begin{eqnarray}
	S(r) = \sum_{i=1,2}\,S_{\rm 0,\,i} \left[1+\left(\frac{r}{r_{\rm c,\,i}}\right)^2 \right]^{-3\beta+0.5} \,.
\end{eqnarray}
Double $\beta$-models provide good representations of cooling-core clusters \citep[e.g.,][]{JonesForman1984,Ota2013}. However, the core of the NE subcluster in MACS~J0416.1-2403 is far from cool, having a temperature $>10$~keV based on the temperature map in Figure~\ref{fig:global_map_fixedZ}. To constrain a possible cooler component, we examined the probability that the gas in bin~\#4 (see Figure~\ref{fig:binmap}) has two phases: one corresponding to a ``hidden'' cool core, and another corresponding to the hot plasma that increases the average core temperature to $>10$~keV. We modelled the spectrum of bin~\#4 with a two temperature APEC model. The normalizations of the two components were free in the fit, as was one of the temperatures; the temperature of the second, cooler thermal component was fixed to $5$~keV. The fit sets an upper limit of $2.82\times 10^{-4}$~cm$^{-5}$~arcmin$^{-2}$ on the normalization of the cooler component (corresponding to a luminosity over $5$ times lower than that of the NE subcluster), and provides no improvement in the statistics of the fit. A cool component with a temperature $<5$~keV would need to be even fainter. Cooler gas in the NE core could be masked by inverse Compton (IC) emission from the active galactic nucleus (AGN) hosted by the NE subcluster brightest cluster galaxy (BCG). To examine this possibility, we measured the temperature in an annulus with radii 32 and 64~kpc around the AGN in the NE BCG. The best-fitting temperature in this annulus is very high, $15.65_{-3.22}^{+5.03}$~keV. In a smaller circle with a radius of 35~kpc around the NE core, the best-fitting temperature is $10.54_{-2.11}^{+3.00}$~keV, and consistent with the temperature calculated in an annulus around the core. Therefore, we find no evidence that the NE core is cool.

From the best-fitting double $\beta$-model, we calculated the central density of the NE core to be $(1.4\pm 0.3)\times 10^{-2}$~cm$^{-3}$. The density and temperature of the NE core hence imply a cooling time of $3.5_{-0.9}^{+1.0}$~Gyr, which further supports our conclusion that the NE subcluster cannot be classified as a cool core cluster based on currently available X-ray data.

\begin{figure*}
	\includegraphics[width=\columnwidth]{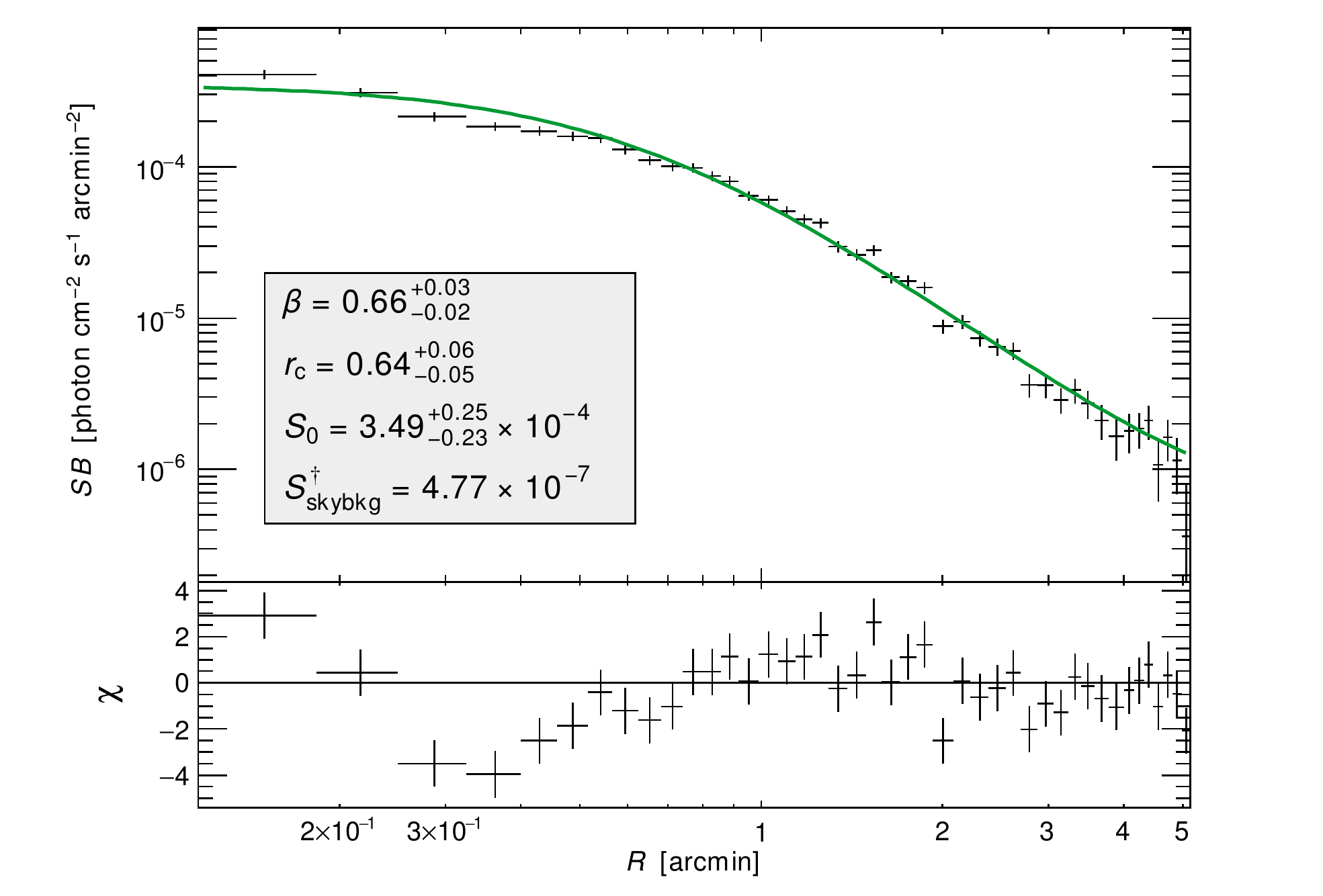}
	\includegraphics[width=\columnwidth]{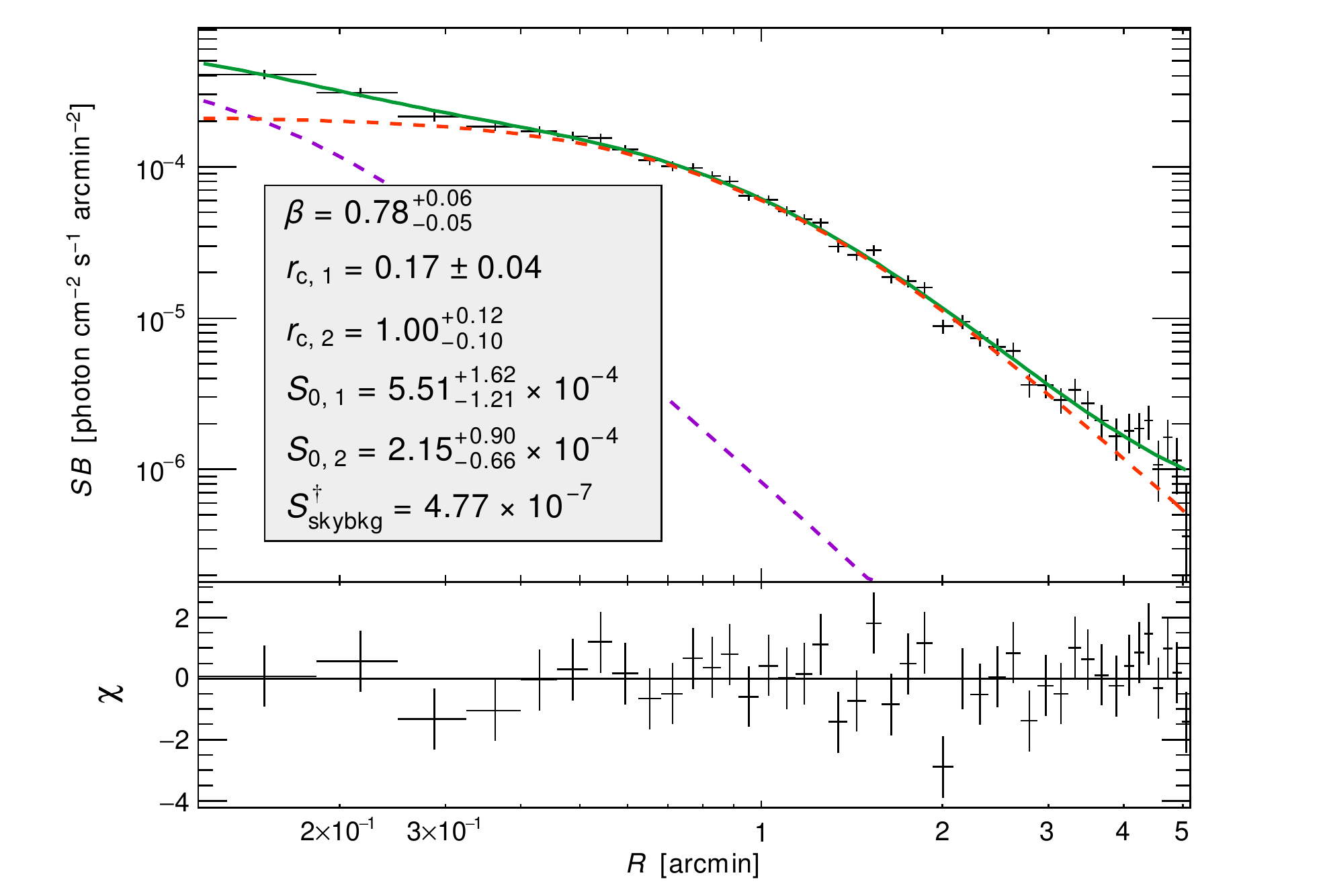}
	\includegraphics[width=\columnwidth]{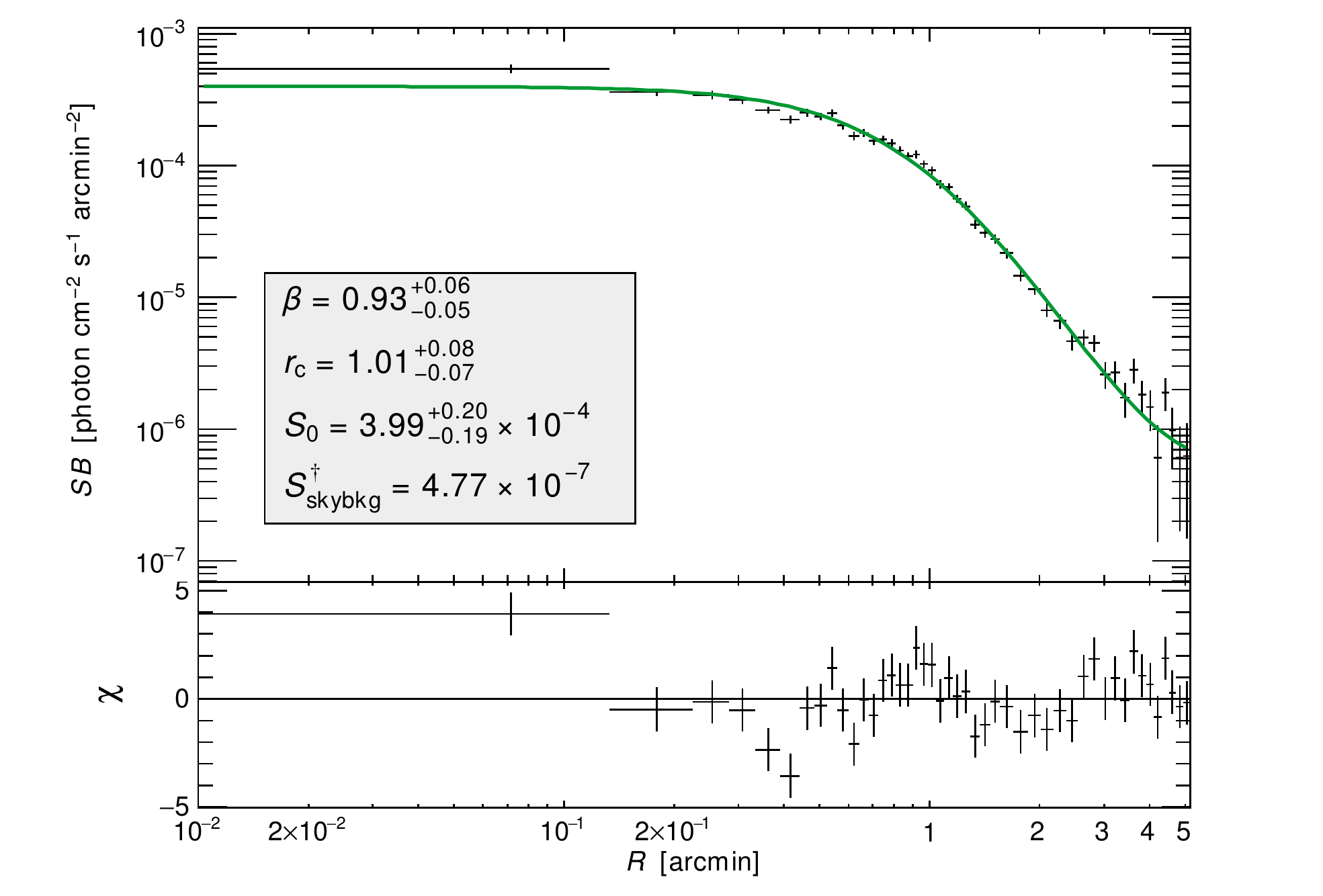}
	\includegraphics[width=\columnwidth]{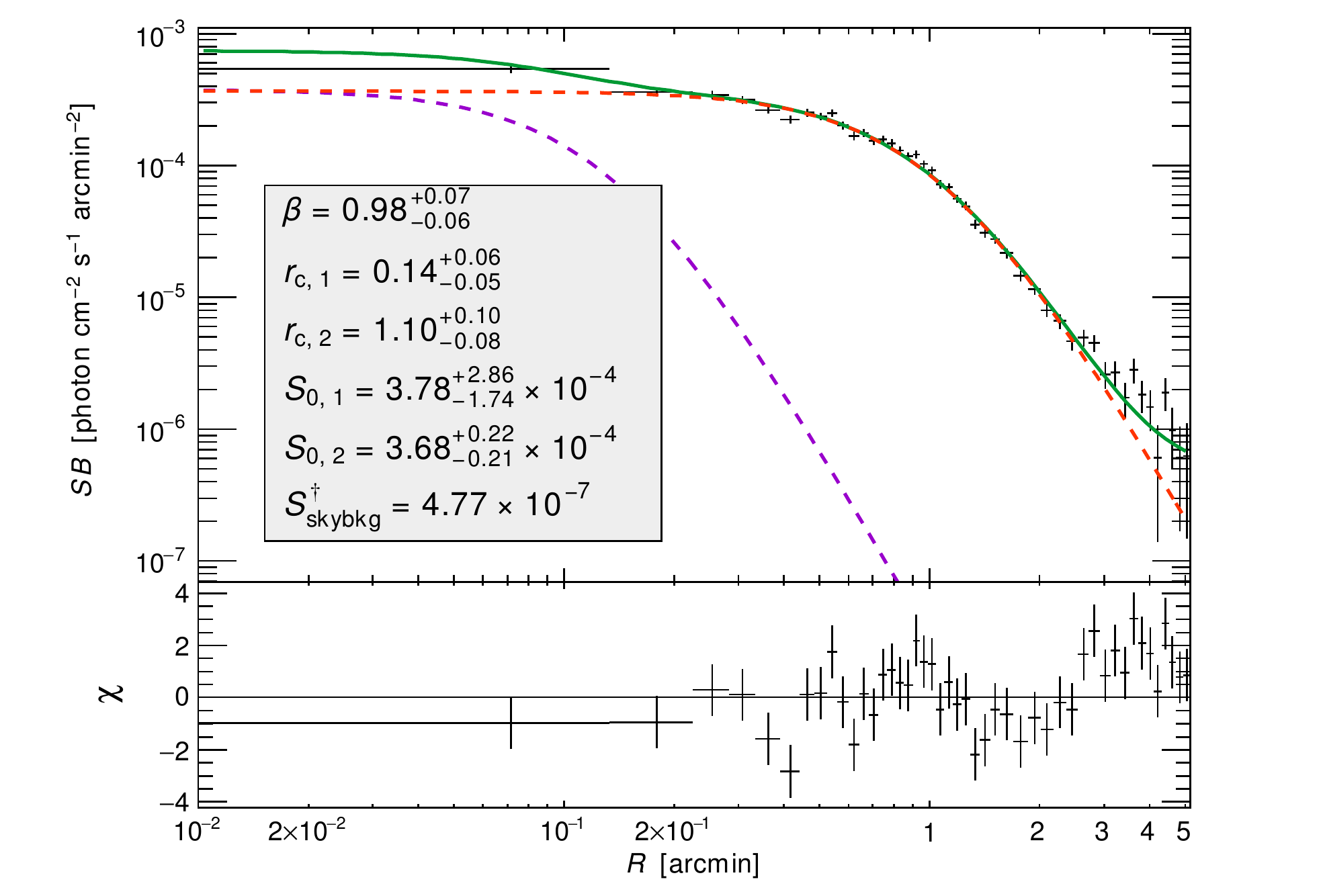}
	\includegraphics[width=\columnwidth]{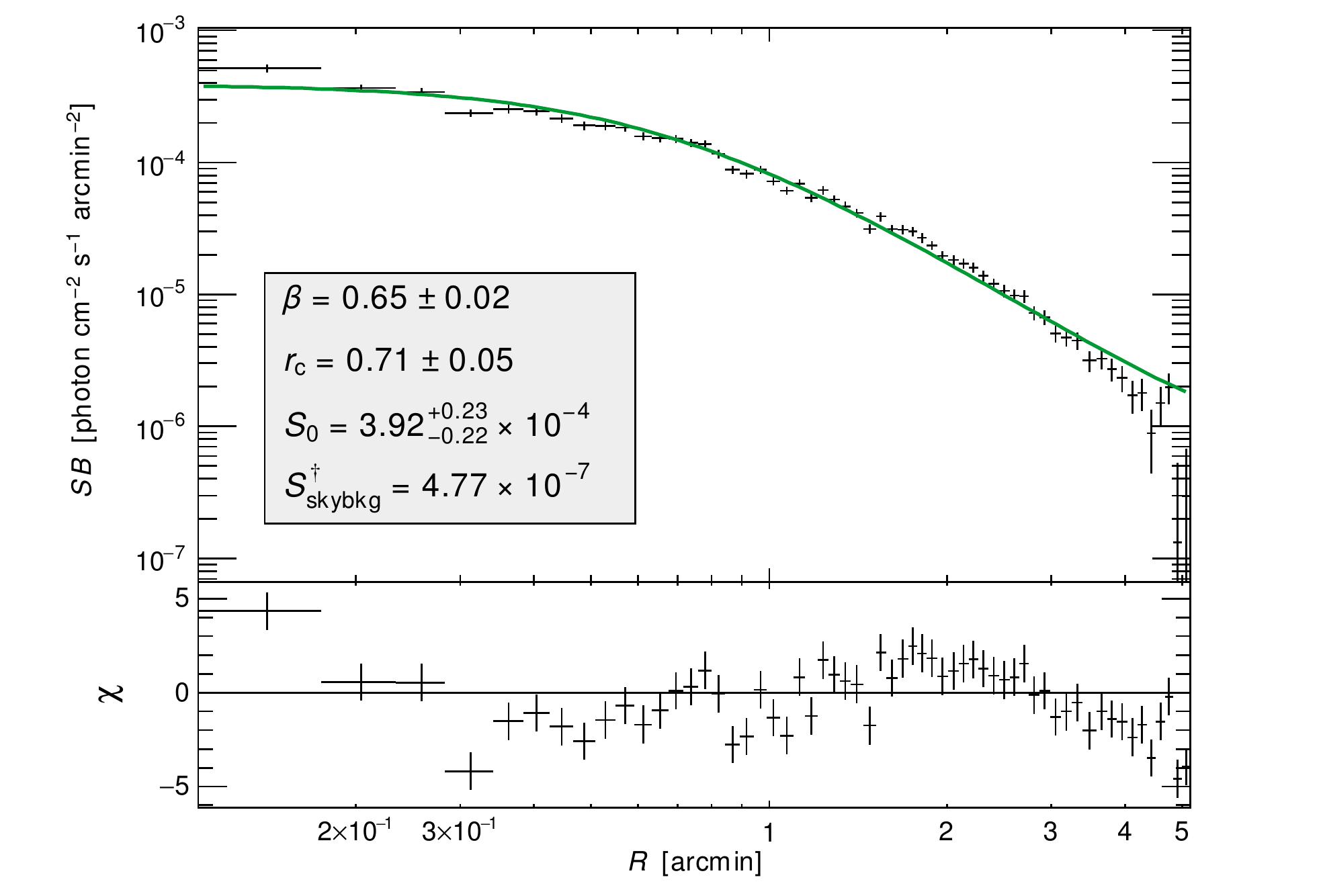}\hfill
	\includegraphics[width=\columnwidth]{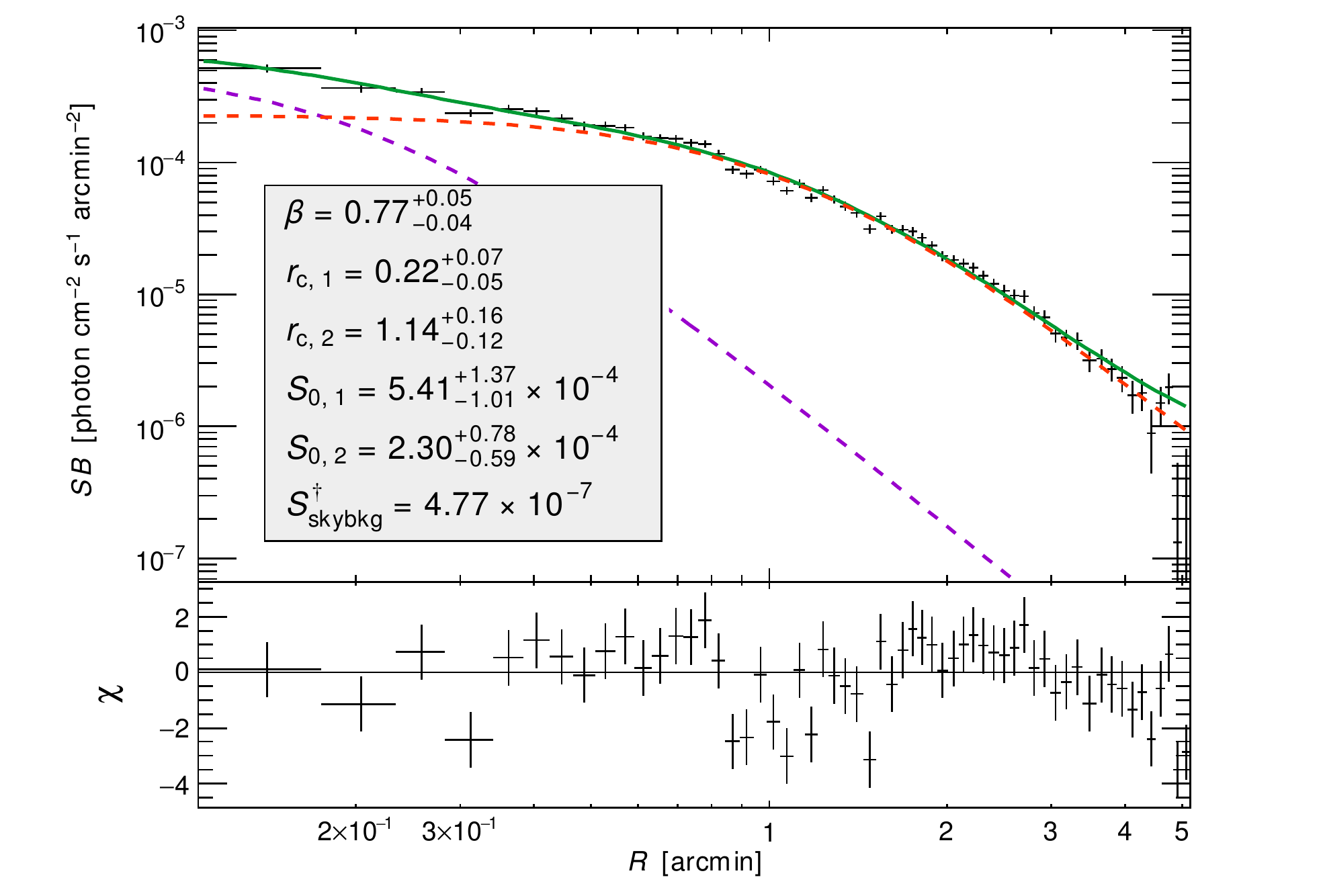}
	\caption{$\beta$-model (left) and double $\beta$-model (right) fits to the surface brightness profiles of the NE sectors with position angles of $45^{\circ}-115^{\circ}$ (top), $115^{\circ}-155^{\circ}$ (middle), and $155^{\circ}-225^{\circ}$ (bottom). The position angles are measured from W, in a counterclockwise direction. For all the profiles, a double $\beta$-model constitutes a better fit than a single $\beta$-model, with confidence levels of  $4.4\sigma$ for the top profile, $3.3\sigma$ for the middle profile, and $6.4\sigma$ for the bottom profile. The individual $\beta$-models are shown with dashed lines. For clarity, all profiles shown have been regrouped in the plots to have 200 counts per bin. The best-fitting model parameters are listed on the plots; radii units are arcmin, and surface brightness units are photon~cm$^{-2}$~s$^{-1}$~arcmin$^{-2}$. Fixed parameters are shown with a superscripted $\dagger$ The bottom panel shows the residuals of the fit.\label{fig:subsectors-N}}
\end{figure*}

To investigate if the double $\beta$-model shape of the NE profile is caused by substructure in a particular direction, we divided the NE sector shown in Figure~\ref{fig:sectors-N} into three subsectors, and modeled the profile of each subsector with $\beta$- and double $\beta$-models. The fits are shown in Figure~\ref{fig:subsectors-N}. For each of the subsectors, a double $\beta$-model describes the profile better than a single $\beta$-model at a confidence level $>99.99\%$.


\subsection{SW Subcluster}

\begin{figure}
	\includegraphics[width=0.9\columnwidth]{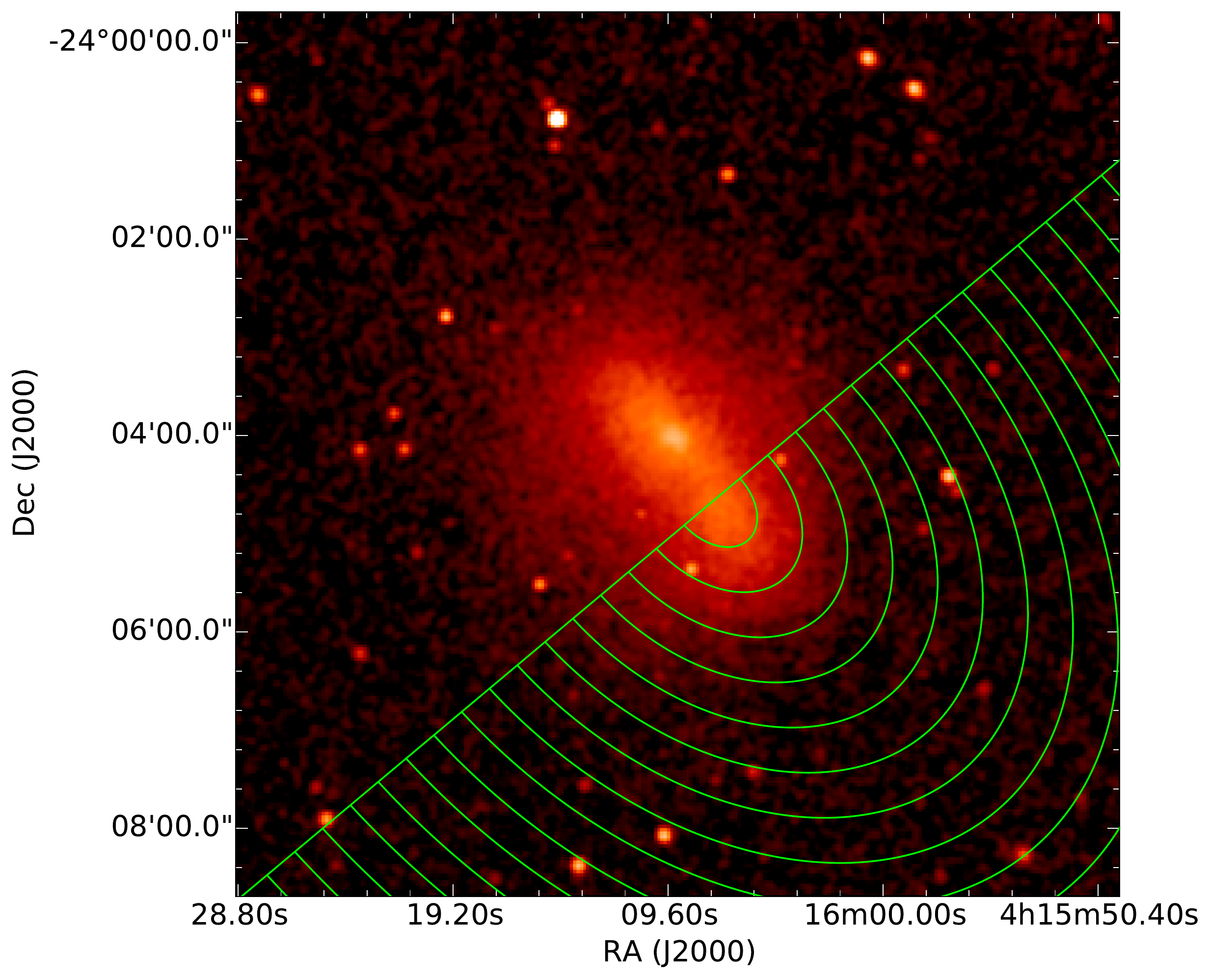}
	\includegraphics[width=\columnwidth]{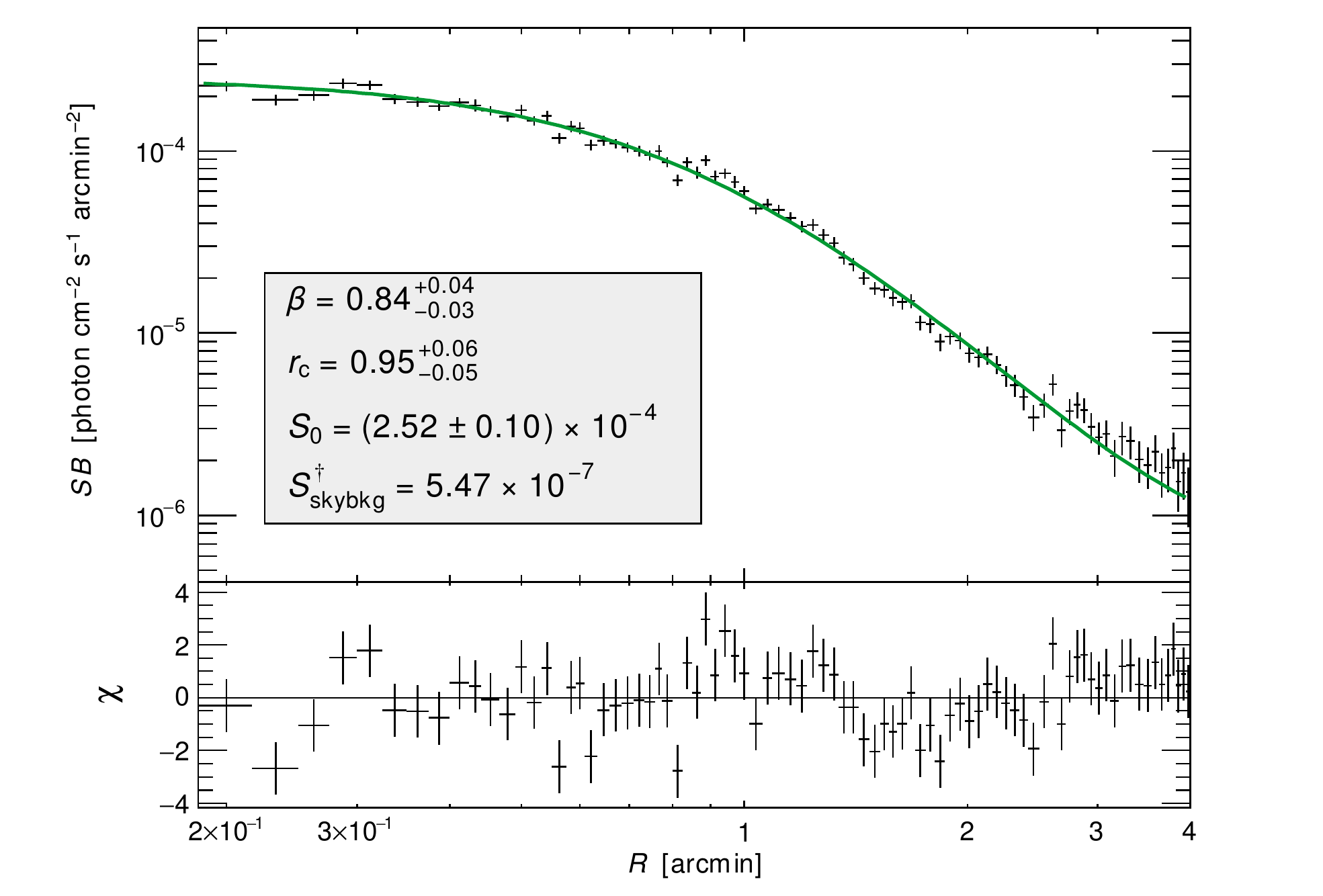}
	\caption{\emph{Top:} Sector used to model the surface brightness of the SW subcluster. Annuli are drawn only to guide the eye and do not reflect the actual bin size used for the surface brightness profiles. \emph{Bottom:} $\beta$-model fit to the surface brightness profile of the SW subcluster. For clarity, the profile shown in this plot was binned to have 200 counts per bin. The best-fitting model parameters are listed; radii units are arcmin, and surface brightness units are photon~cm$^{-2}$~s$^{-1}$~arcmin$^{-2}$. Fixed parameters are shown with a superscripted $\dagger$. The bottom panel shows the residuals of the fit. \label{fig:sectors-S}}
\end{figure}

The surface brightness profile of the SW subcluster and the sector used in the extraction of this profile are shown in Figure~\ref{fig:sectors-S}. As for the NE profiles, the sky background profile was modelled by fitting a constant to the outer bins of the profile, in the range $r=4.5\arcmin-10.7\arcmin$. Unlike the NE profiles, the SW profile is modelled well by a simple $\beta$-model. The fit is shown in Figure~\ref{fig:sectors-S}. 

\begin{figure}
	\includegraphics[width=\columnwidth]{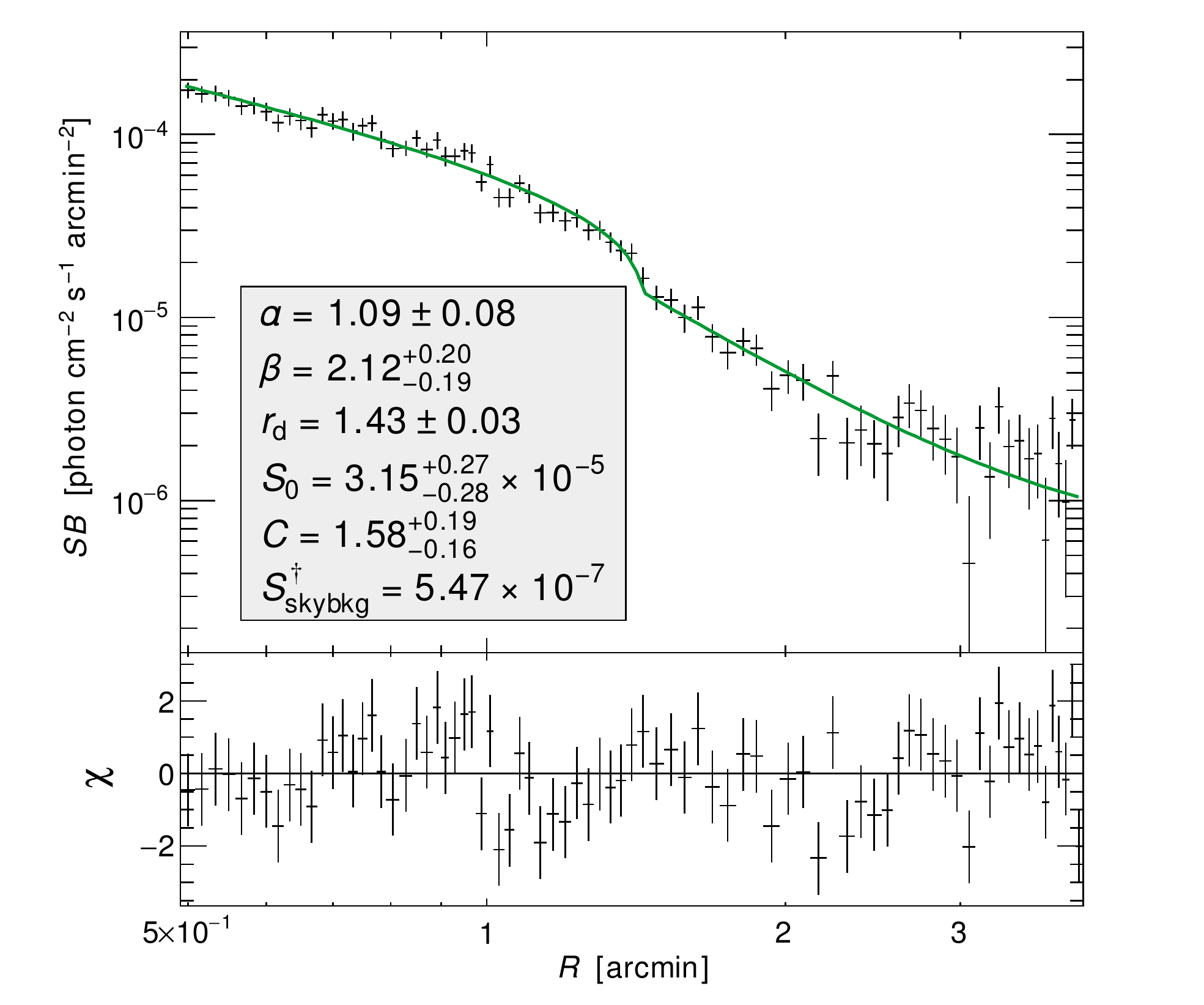}
	\caption{Broken power-law model fit to the surface brightness profile of the SW subcluster in a subsector with position angles $280^{\circ}-340^{\circ}$, measured from the W in counterclockwise direction. For clarity, the profile shown in this plot was binned to have 80 counts per bin. The best-fitting model parameters are listed; radii units are arcmin, and surface brightness units are photon~cm$^{-2}$~s$^{-1}$~arcmin$^{-2}$. Fixed parameters are shown with a superscripted $\dagger$.\label{fig:subsector-S}}
\end{figure}

A weak edge in the profile can be seen at $r\sim 1.5\arcmin$. We divided the profile into three subsectors with equal opening angles ($60^{\circ}$) to check whether the edge is seen along all three directions, and found that it is present only in the central subsector (position angles $280^{\circ}-340^{\circ}$, measured from the W in a counterclockwise direction). To describe it, we fitted a projected broken power-law elliptical density model to the surface brightness profile between $r=0.5\arcmin-4.0\arcmin$. The density model is defined as:
\begin{eqnarray}
	n(r) = \begin{cases} C\,n_{\rm {0}} \left(\frac{r}{r_{\rm d}}\right)^{-\alpha}\,, & \mbox{if } r \le r_{\rm d} \\ n_{\rm {0}} \left(\frac{r}{r_{\rm d}}\right)^{-\beta}\,, & \mbox{if } r > r_{\rm d} \end{cases} \,,
\end{eqnarray}
where $n$ is the electron number density, $C$ is the density compression, and $r_{\rm d}$ is the radius of the density jump. The fit is shown in Figure~\ref{fig:subsector-S}. If the density discontinuity is a shock front, then its magnitude corresponds to a shock with Mach number $\mathcal{M}=1.40_{-0.12}^{+0.14}$. Unfortunately, the count statistics are too poor to allow us to distinguish between a cold front and a shock front based on the temperature jump, and thus we cannot determine the nature of the surface brightness edge.

\section{Substructure in the ICM}
\label{sec:substructure}

\begin{figure}
	\includegraphics[width=\columnwidth]{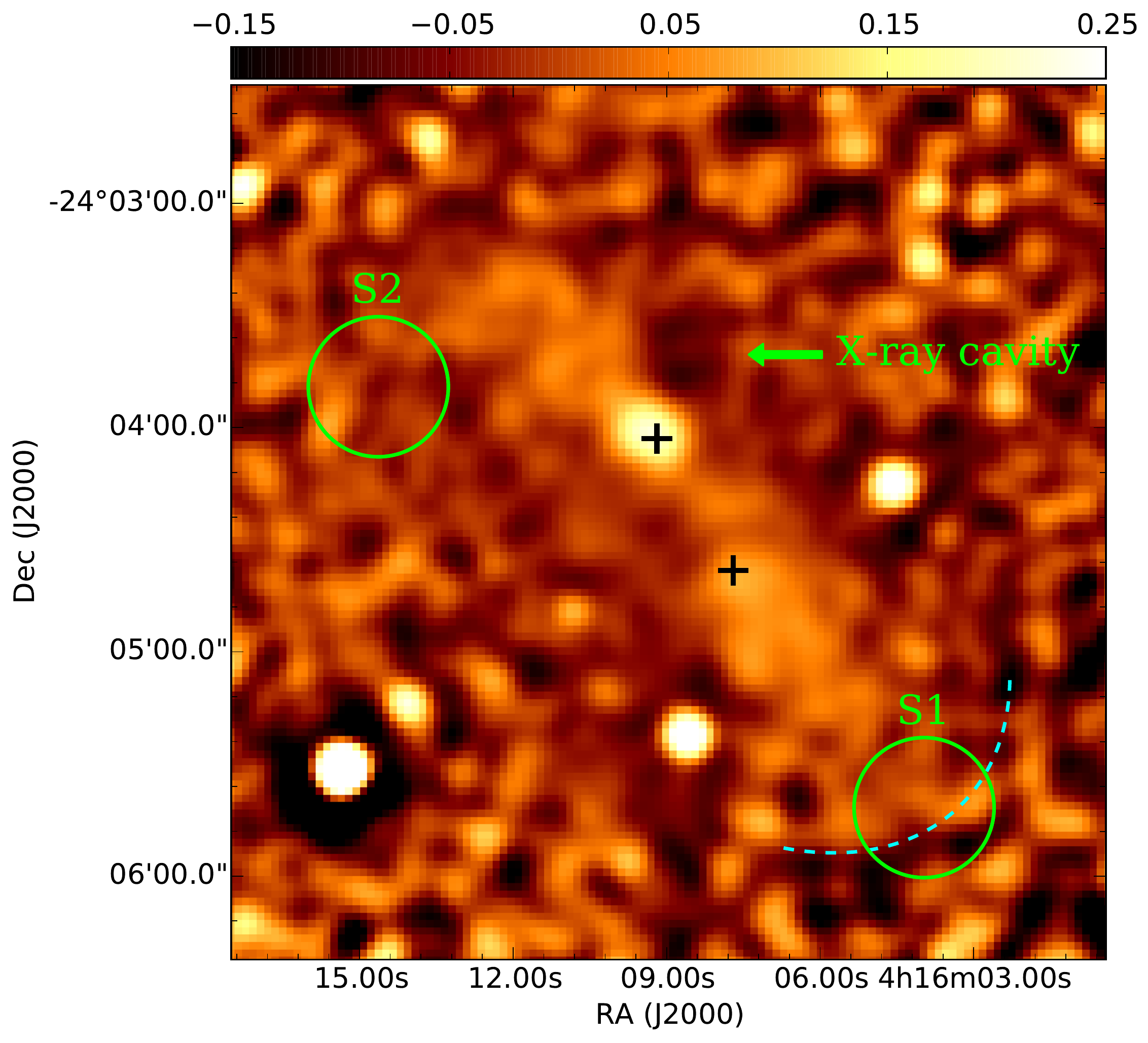}
	\caption{Unsharp-masked X-ray image of MACS~J0416.1-2403, created by dividing the difference of two $0.5-4$~keV fluxed images convolved with Gaussians of widths $4\arcsec$ and $10\arcsec$ by their sum. Point sources have not been subtracted from this image, and there are some residuals surrounding them. The positions of the two less massive mass structures identified by \citet{Jauzac2015} are marked as S1 and S2 with circles of diameters $100$~kpc, as also done by \citet{Jauzac2015}. Black crosses mark the centers of the DM halos of the two main subclusters (M. Jauzac, priv. comm.). The dashed arc shows the position of the density discontinuity detected near the SW subcluster. The location of the X-ray cavity is also marked. \label{fig:unsharp}}
\end{figure}

The search for substructure in the ICM is motivated by the identification in the lensing maps of two less massive structures in addition to the main NE and SW subclusters \citep[][]{Jauzac2015}. The positions of these mass structures are shown in Figure~\ref{fig:unsharp} and denoted by S1 and S2 for consistency with the notation of \citet{Jauzac2015}. In the analysis of \citet{Jauzac2015}, S1 and S2 were found to be X-ray-dark; however, the X-ray data presented here are $\sim 6$ times deeper, which would make it easier to observe ICM substructure. 

We searched for substructure using the unsharp-masked image of the cluster, which was created by dividing the difference of two $0.5-4$~keV fluxed images convolved with Gaussians of widths $4\arcsec$ and $10\arcsec$ by their sum. The resulting image, shown in Figure~\ref{fig:unsharp}, highlights substructure on scales of $\sim 20-50$~kpc.\footnote{Smoothing with Gaussians of larger widths does not reveal additional substructure.} There is no excess X-ray emission at the positions of S1 and S2. However, the emission is elongated in the direction of both mass structures. In the south, the ICM appears elongated in the direction of S1, while in the north the emission is elongated along the line connecting the NE and SW subclusters and then appears to curve in the direction of S2. We note that if S1 and S2 have already merged with the NE and SW subclusters, the dark matter would have decoupled from the gas, and therefore we do not necessarily expect a spatial overlap between the dark matter and gas components.

\begin{figure*}
   \includegraphics[height=7cm]{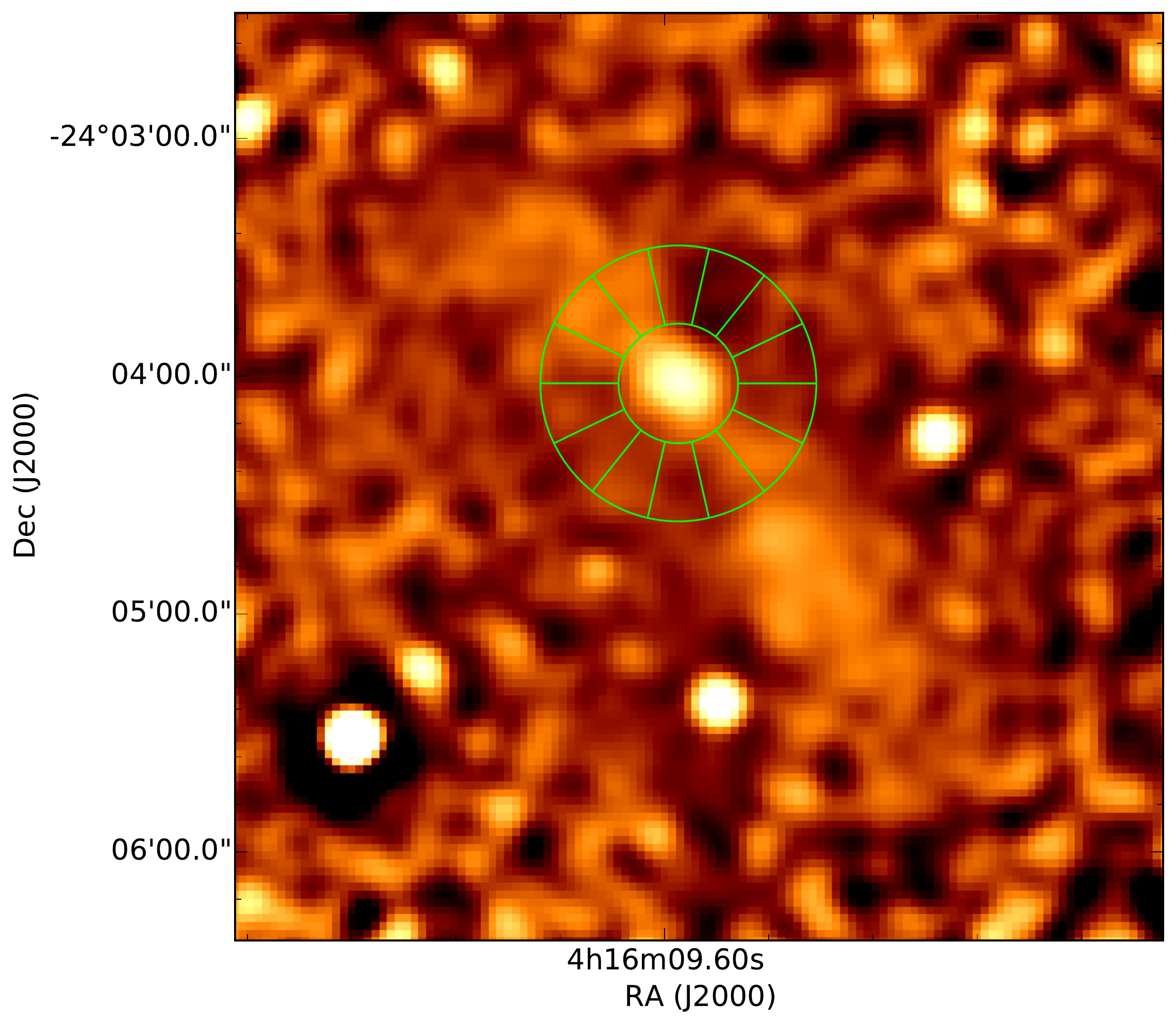}
   \includegraphics[height=7.6cm]{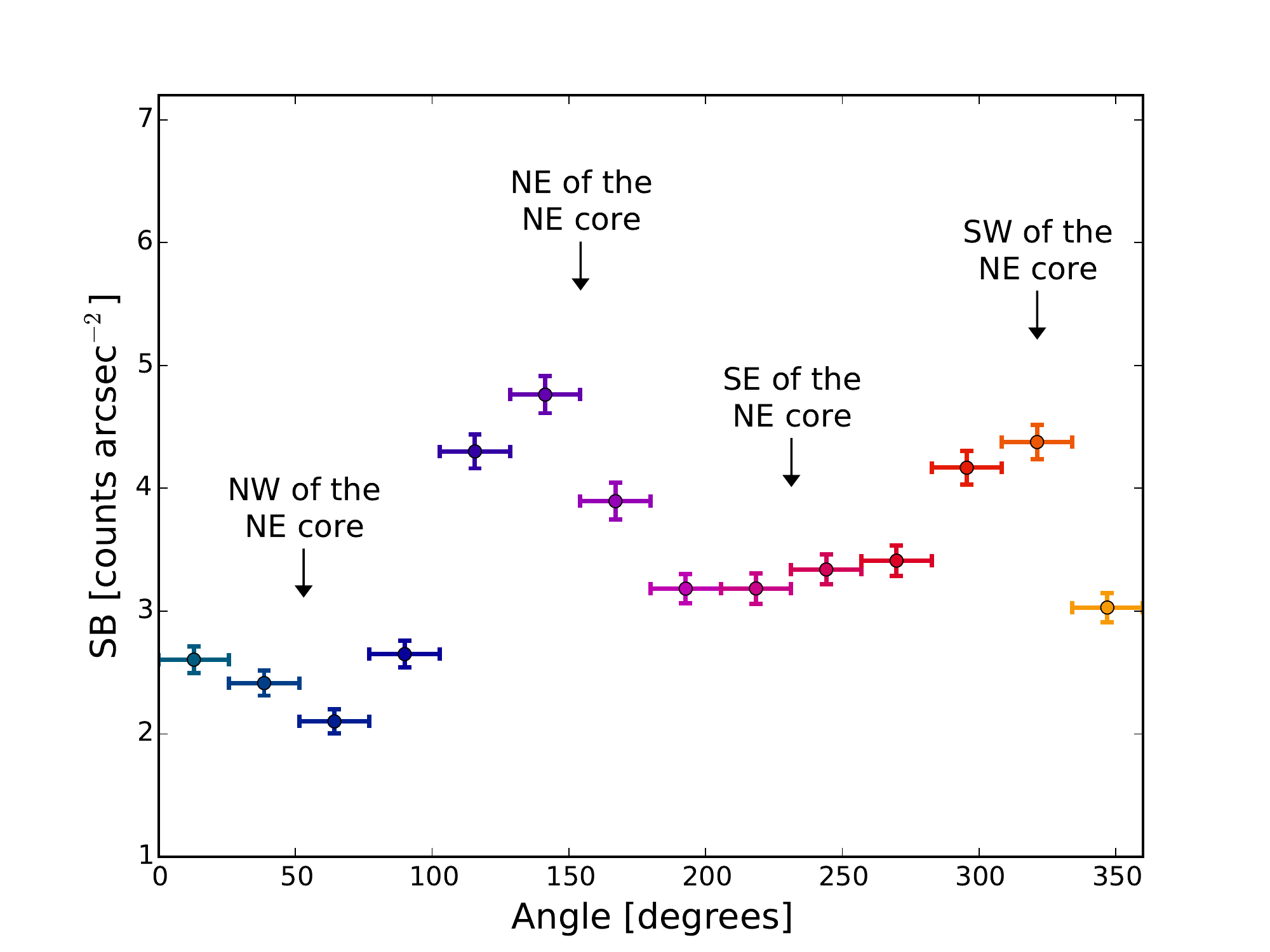}
   \caption{\emph{Left:} Unsharp-masked image as in Figure~\ref{fig:unsharp}, with overlaid partial annuli used to evaluate the azimuthal surface brighness profile around the NE core. \emph{Right:} Azimuthal surface brightness profile around the NE core. The largest dip in the profile is in the direction of the cavity seen in the unsharp-masked image, between opening angles of $51-77$~degrees. Angles are measured counterclockwise, starting from the W. \label{fig:bubble}}
\end{figure*}

Interestingly, the unsharp-masked image enchances a small cavity with a diameter $\sim 50$~kpc NW of the core of the NE subcluster. We examined the significance of the cavity from the azimuthal surface brightness profile in an annulus around the cluster center. The annulus was divided in 14 partial annuli with equal opening angles. The partial annuli and the azimuthal surface brightness profile are shown in Figure~\ref{fig:bubble}. The azimuthal surface brightness profile is lowest in 4 partial annuli located NW of the NE core. The largest dip is in the partial annulus that crosses the middle of the X-ray cavity seen in the unsharp-masked image; the opening angles of this partial annulus are $51-77$~degrees. No radio emission fills the X-ray cavity, but the cavity was likely inflated by the AGN hosted by the NE brightest cluster galaxy (see Section~\ref{sec:jvla-results}). 

\section{Radio results}
\label{sec:jvla-results}

\begin{figure*}
	\includegraphics[width=\columnwidth]{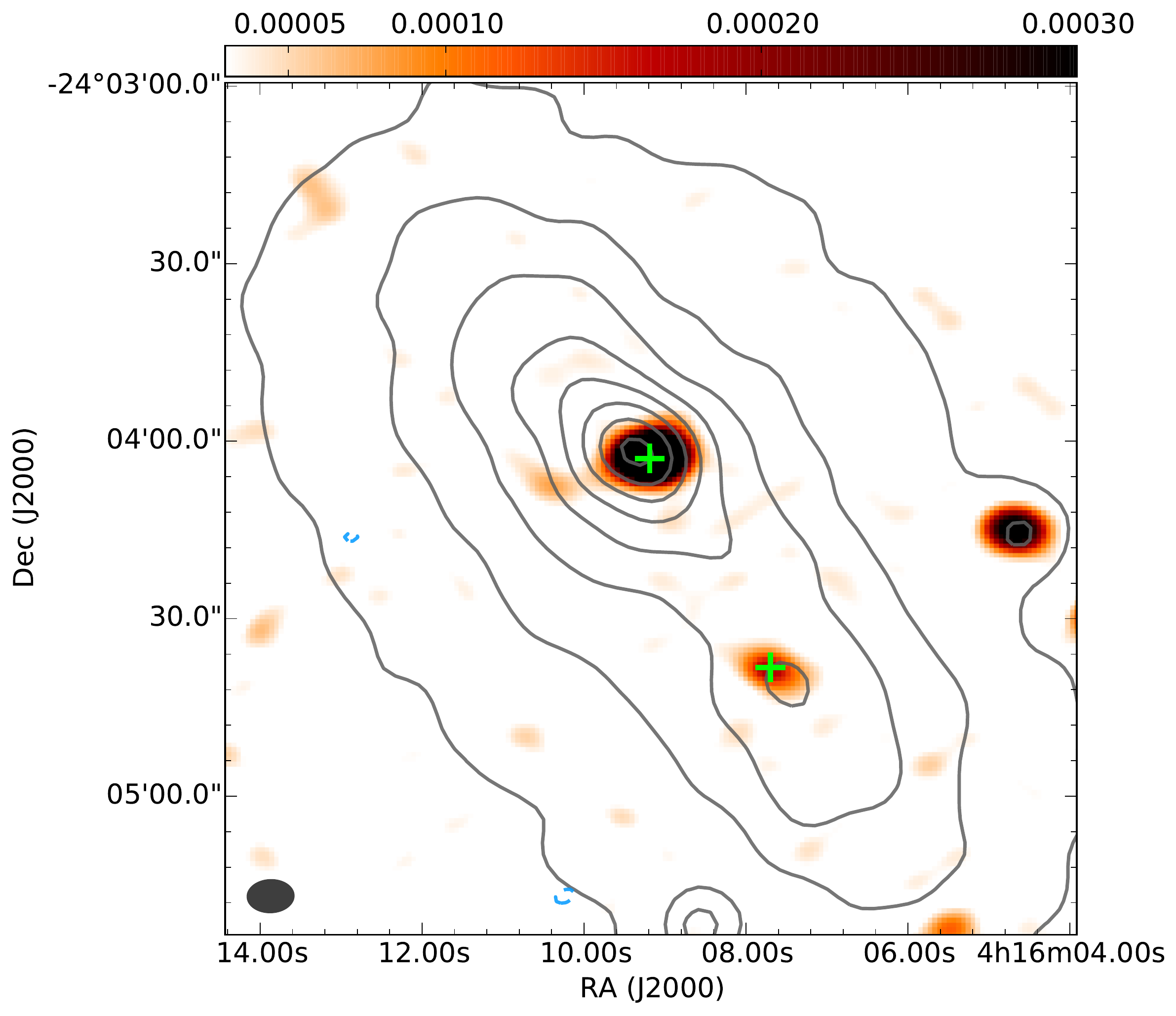}\hfill
	\includegraphics[width=\columnwidth]{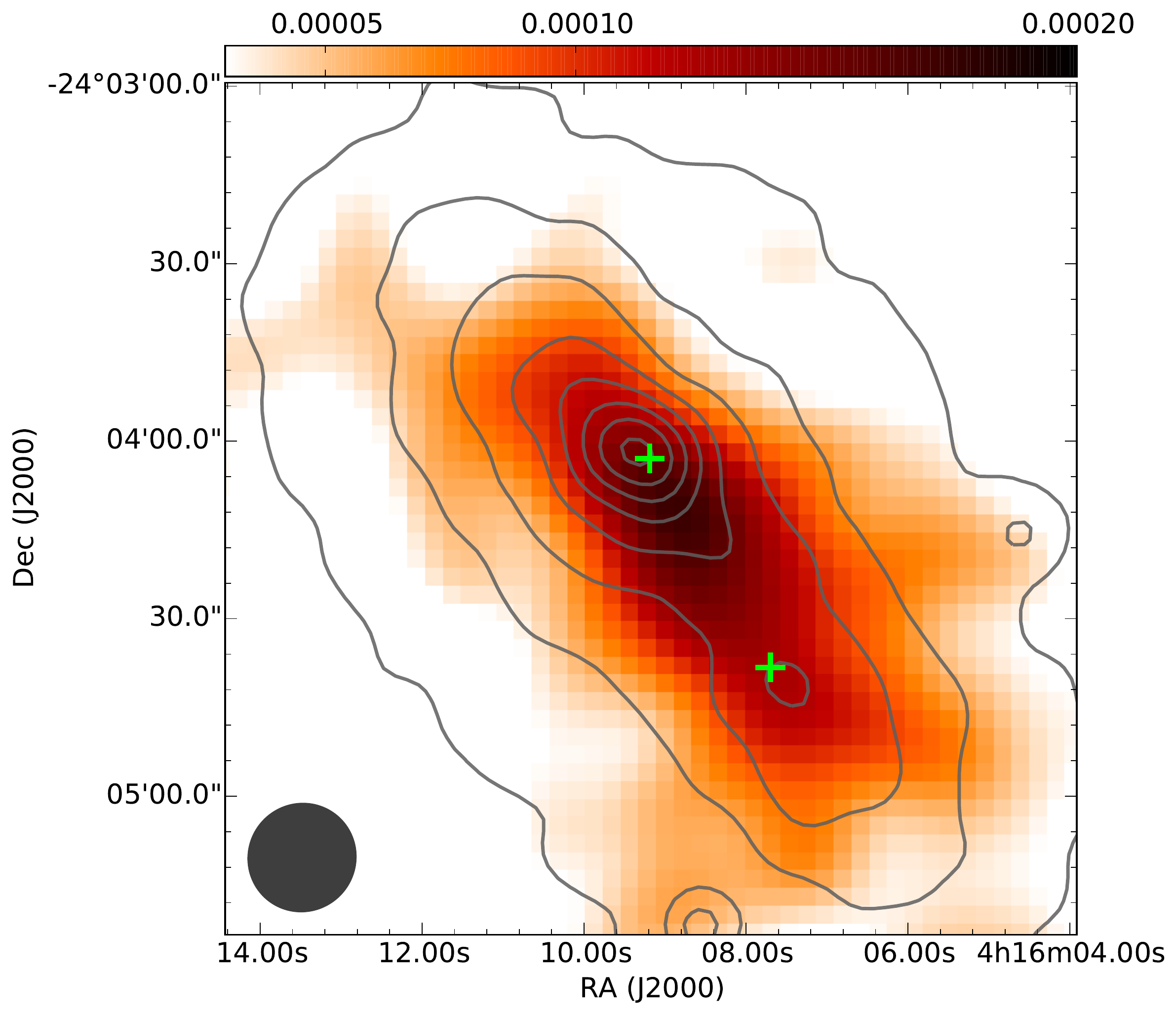}
	\caption{\emph{Left:} JVLA $1-2$~GHz high-resolution image showing the compact sources in the cluster 
	region. \emph{Right:} JVLA $1-2$~GHz image of the radio halo, with compact sources 
	subtracted. \chandra\ contours are overlaid on both images. The contours are based on an 
	exposure-corrected, vignetting-corrected image that was smoothed with a Gaussian kernel of 
	width $4\arcsec$; they are drawn at $[0.6,1.3,2.0,2.7,...]\times 10^{-7}$ photons~cm$^{-2}$~s$^{-1}$. 
	Green crosses mark the centers of the DM halos. Dashed blue lines show the $-3\sigma$ radio 
	contours (these are only visible in the high-resolution radio image). The beam size is shown in the bottom left corner of the images.\label{fig:radio}}
\end{figure*}

\begin{figure}
        \includegraphics[width=\columnwidth]{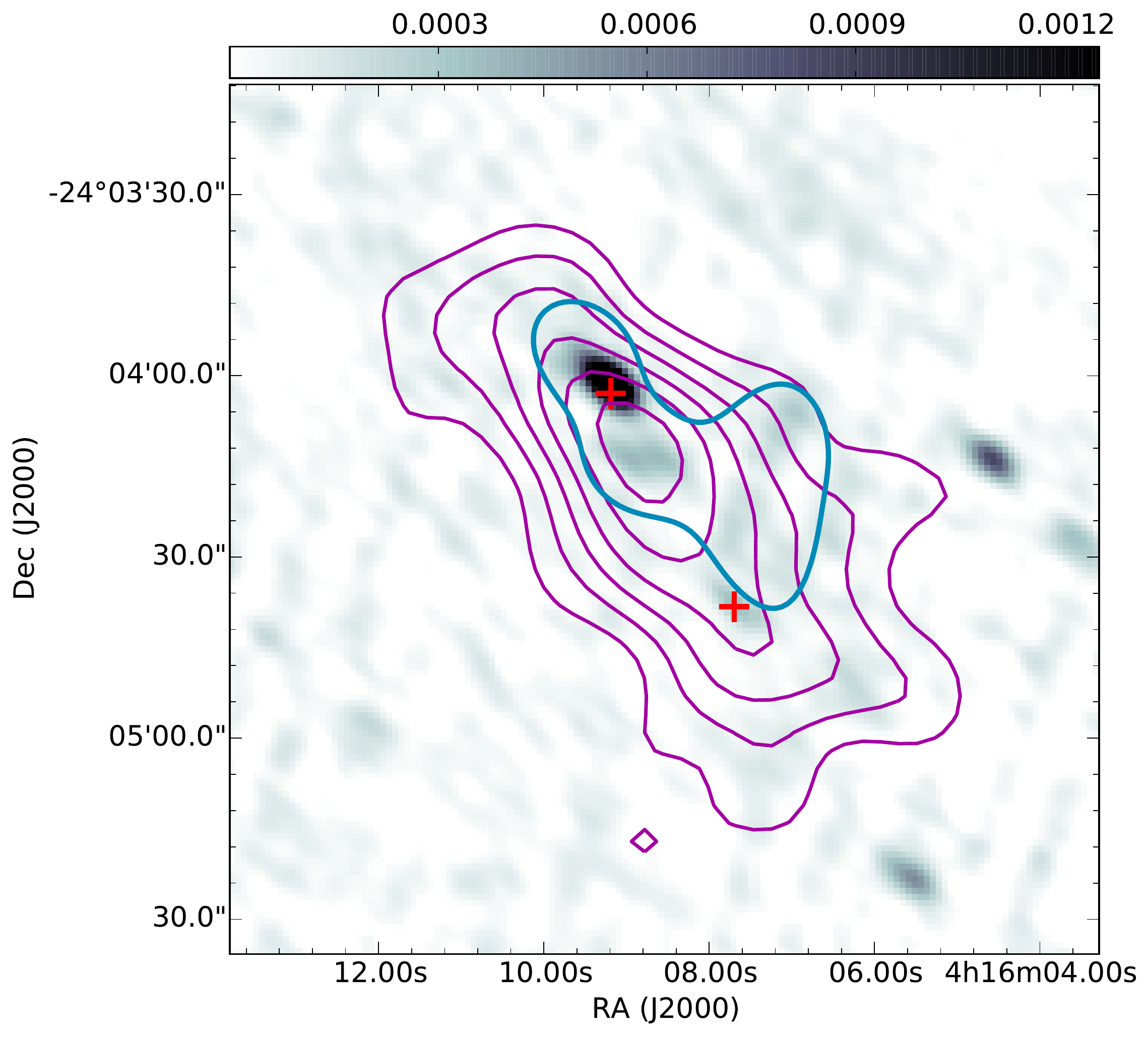}
        \caption{GMRT 610~MHz high-resolution image, showing the compact radio sources. Overlaid in blue are GMRT 610~MHz 
		contours of the diffuse radio emission, drawn at $1.1\times 10^{3}$~$\mu$Jy/beam. 
		JVLA $1-2$~GHz radio contours showing diffuse radio emission are drawn in magenta  
		at $[60,80,100,120,140,160]$~$\mu$Jy/beam. \label{fig:gmrt}}
\vspace{0.45cm}
\end{figure}

The JVLA $1-2$~GHz images reveal several compact sources in the cluster region (Figure~\ref{fig:radio}). 
Two of these sources are associated with cD galaxies in the NE and SW subclusters. 
These sources are also detected in the GMRT 610~MHz image (Figure~\ref{fig:gmrt}).
These two point-like AGN have $1.5$~GHz integrated flux 
densities\footnote{The quoted flux errors include a 5\% uncertainty for the absolute flux scale bootstrapped 
from the primary calibrator sources. For the GMRT observations, we assume an uncertainty 
in the absolute flux scale of 10\%.} of $1.47 \pm 0.08$ (NE) and $0.27 \pm 0.03$ (SW) mJy.
At 610~MHz we measure flux densities of $3.24\pm0.34$ (NE) and $0.33 \pm 0.07$ (SW) mJy for these sources.

We also find diffuse extended emission in the cluster. The diffuse emission reveals itself in the JVLA 
image as an increase in the ``noise''  in the general cluster area. 
This diffuse emission is better visible in our low-resolution tapered image with 
emission from compact sources subtracted (Figure~\ref{fig:radio}). The diffuse 
emission has a elongated shape measuring about $120\arcsec\times45\arcsec$ ($0.65$ by $0.24$~Mpc) and 
is oriented along a NE-SW axis, following the overall distribution of the X-ray emission. 
We also find evidence for this diffuse emission in the GMRT 610~MHz image, 
although less clearly than in the JVLA 1.5~GHz image (Figure~\ref{fig:gmrt}). In the low-resolution tapered GMRT image there 
is again evidence for diffuse emission, but the peak flux is only at a level of $3\sigma_{\rm{rms}}$.

For the integrated flux of the diffuse emission we measure $1.58 \pm 0.13$~mJy (at 1.5 GHz) 
in an ellipse with radii of 1\arcmin\ by 0.5\arcmin\ orientated with a position angle of 
45\degr\ (following the overall brightness distribution). 
From the GMRT tapered image we estimate an integrated flux of $6.8\pm3.0$~mJy for the diffuse emission in the same area. 

Taking the two flux measurements at 1.5 and 0.61~GHz, we obtain a spectral 
index of $\alpha_{1500}^{610} = -1.6\pm 0.5$ for the diffuse emission. Based on the above results, we compute a radio power of 
$P_{1.4\,\,\rm{GHz}} = (1.06 \pm 0.09) \times 10^{24}$~W~Hz$^{-1}$, 
scaling with the spectral index of $\alpha = -1.6$.

VLITE shows emission co-incident with the NE
subcluster where the higher resolution $1-2$~GHz images
(Figure~\ref{fig:radio}) reveal the compact sources and the brightest
portion of the diffuse emission. We fit the 320--360~MHz VLITE emission with a
single Gaussian component to measure the integrated flux. We measure a
total flux of $17 \pm 6$~mJy where we have included an 8\%
uncertainty in the absolute flux scale. The VLITE emission contains
both the diffuse component and the compact emission from the two
point sources associated with the NE and SW clusters. We use the VLA and
GMRT flux measurements to determine the spectral index of the two compact
sources, assume that spectral index extends to the central VLITE frequency, 
and estimate the contribution to the total VLITE emission from the compact
sources. We subtract that from the VLITE flux and get an estimate of the diffuse
component detected by VLITE of $11 \pm 4 $~mJy. Comparing the VLITE flux to the 
JVLA flux, we calculate a spectral index of $\alpha_{1500}^{340} = -1.3 \pm 0.3$. The results are 
consistent with the spectral index estimate from the GMRT data. Deeper
low-frequency data are required to better constrain the spectral properties of the
diffuse emission.

In the JVLA $1-2$~GHz data we did not detect diffuse polarized
emission in the cluster in Stokes Q and U images we made. Halos are generally 
unpolarized at the few percent level or less so this result is not surprising \citep[e.g.,][]{Feretti2012}. 
We note however, that a proper search for diffuse emission, taking full
advantage of the large bandwidth, would require Faraday Rotation
Measure Synthesis.


\section{Gas-DM Decoupling}
\label{sec:decoupling}

The combined X-ray and lensing analysis of \citet{Jauzac2015} determined that the gas component of the SW subcluster has decoupled from the DM component and lags behind the subcluster's galaxies as it travels towards the NE subcluster. In the NE subcluster, on the other hand, the DM and gas components were determined by \citet{Jauzac2015} to spatially overlap.\footnote{An analysis of the DM-gas offset in MACS~J0416.1-2403 was carried out more recently by \citet{Harvey2015}. For the SW subcluster, their DM peak is offset by about 100~kpc ($\sim 20\arcsec$) from the DM peak determined by previous analyses \citep[e.g.,][models at https://archive.stsci.edu/pub/hlsp/frontier/macs0416/models/]{Jauzac2015,Grillo2015}. We do not understand the cause of the offset, but choose to use the DM peaks determined by \citet{Jauzac2015}, because their locations are consistent within $2\arcsec$ with the locations of the DM peaks determined by other authors \citep[e.g.,][]{Grillo2015}. We also point out that the peak of the galaxy distribution chosen by \citet{Harvey2015} for the SW subcluster in MACS~J0416.1-2403 is strongly biased by a foreground galaxy \citep[$z=0.10_{-0.08}^{+0.12}$;][]{Jouvel2014} at ${\rm RA}=04:16:06.833$ and ${\rm DEC}=-24:05:08.40$ that was not excluded from their analysis.} Offsets between the DM and gas components of merging galaxy clusters set crucial constraints on the merger state. In particular, significant offsets indicate a post-merging system, while a lack of offsets supports a pre-merging scenario. Below, we discuss the DM-gas offsets in light of the deeper \chandra\ data. 

\begin{figure*}
	\includegraphics[width=\columnwidth]{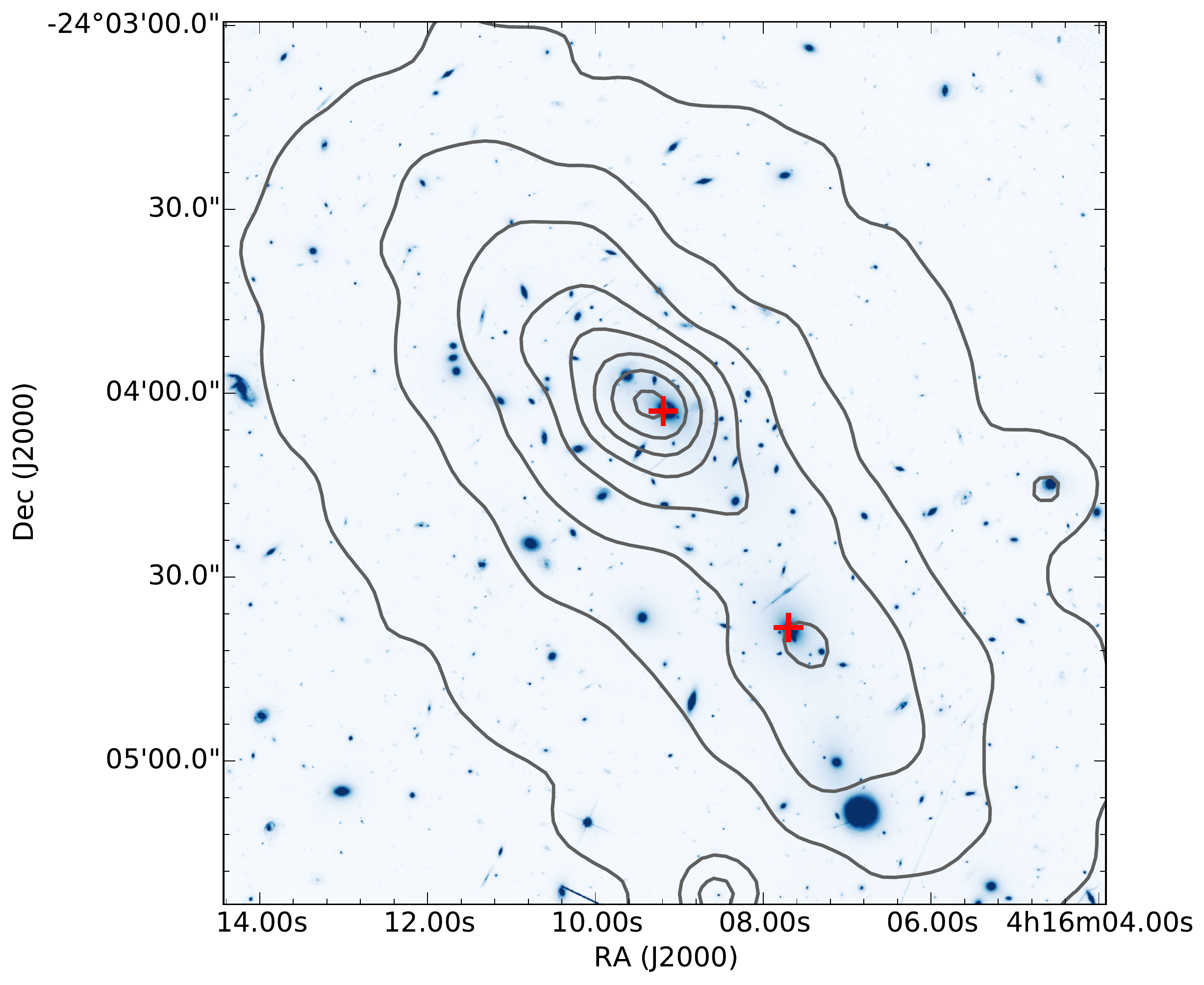}\hfill
	\includegraphics[width=\columnwidth]{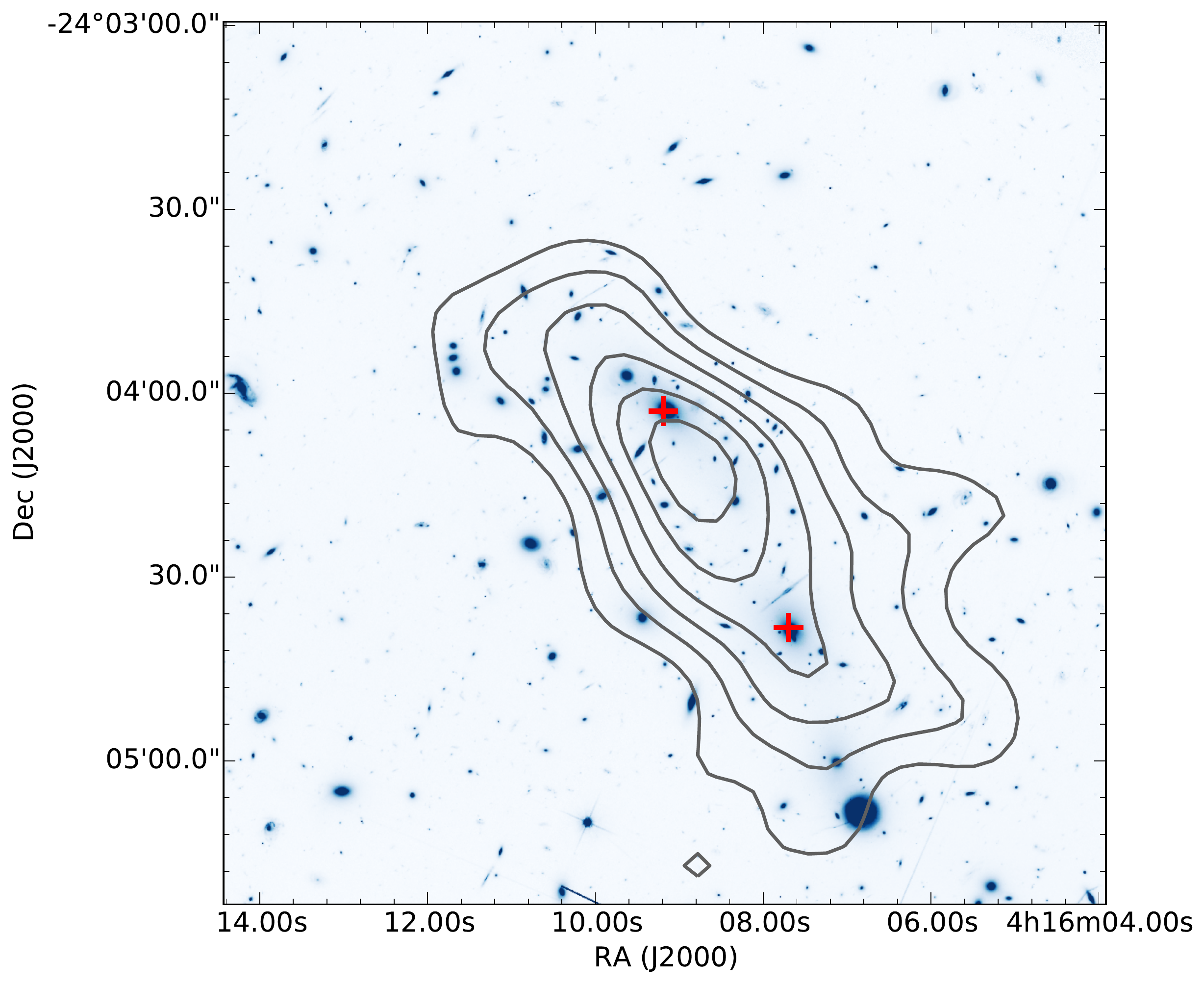}
	\caption{HST ACS/F814W optical image of MACS~J0416, with overlaid \chandra\ (left) and JVLA (right) contours. X-ray contours are the same as in Figure~\ref{fig:radio}. Radio contours are based on the JVLA image with the emission from compact sources subtracted using a Gaussian uv-taper of 20\arcsec. The JVLA contours are drawn at $[60,80,100,120,140,160]$~$\mu$J/beam. The centers of the DM halos of the NE and SW subclusters are marked with red crosses. \label{fig:optical-xray}}
\end{figure*}

\begin{figure*}
	\includegraphics[width=\columnwidth]{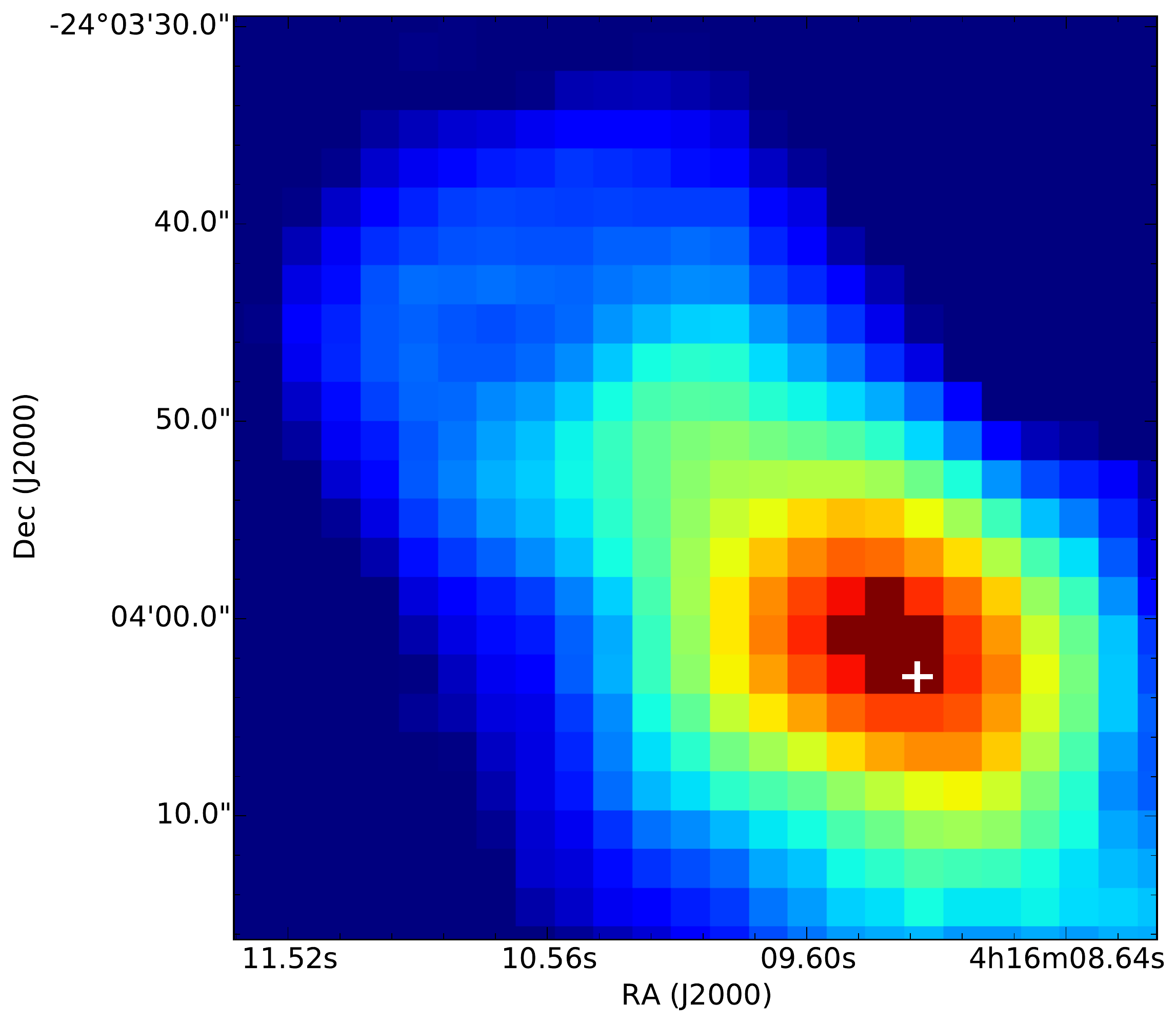}\hfill
	\includegraphics[width=\columnwidth]{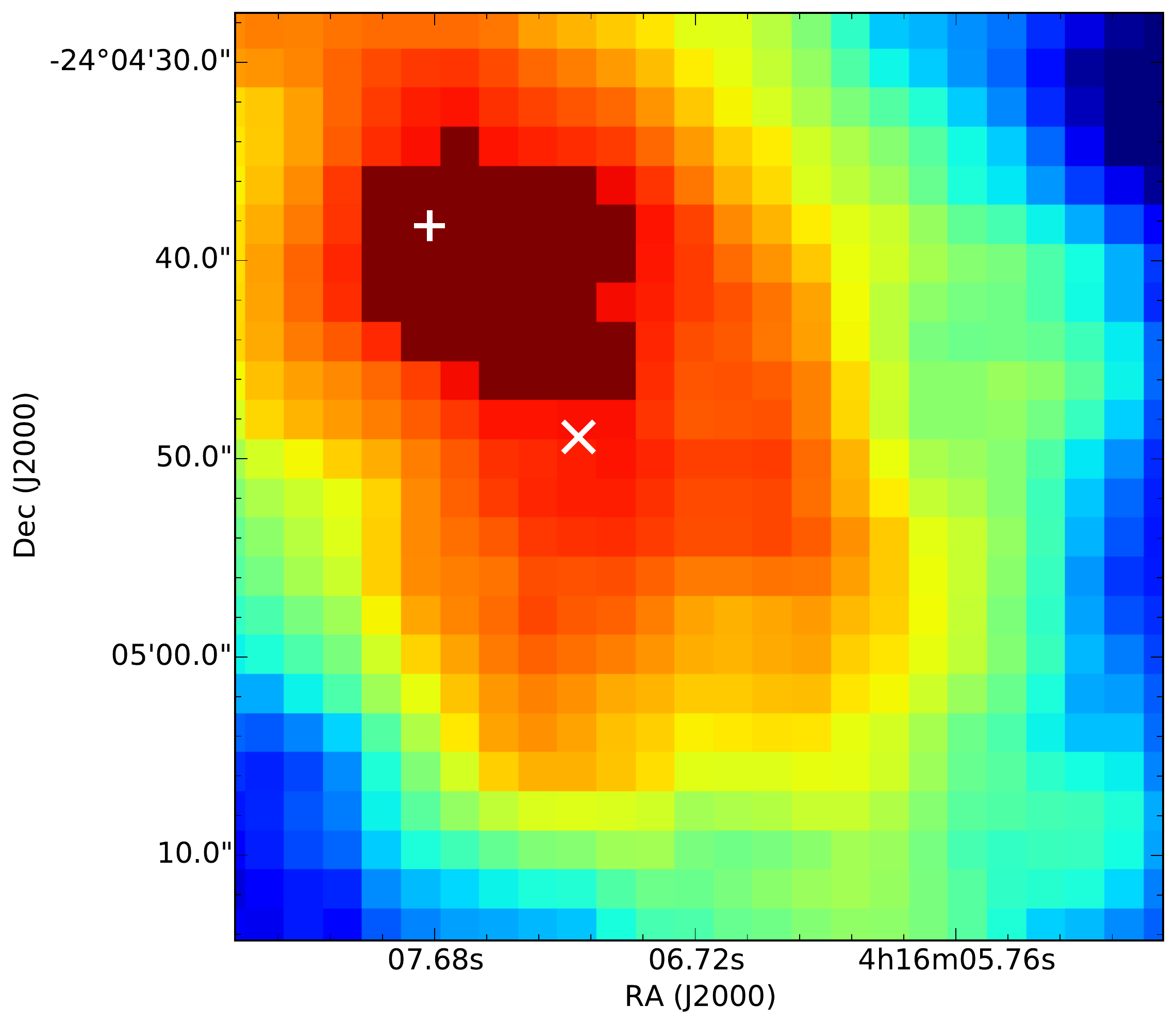}
	\caption{\chandra\ $0.5-4$~keV surface brightness maps of the NE subcluster center (left) and of the SW subcluster center (right). The overexposed regions (in brown) are the peak regions of the two subcluster cores. The position of the DM centers are marked with white crosses. The white `X' indicates the location of the X-ray peak of the SW subcluster, as it was identified from shallower \chandra\ data by \citet{Jauzac2015}. \label{fig:DM-zoom}}
\end{figure*}

The centers of the DM halos of the two subclusters (M. Jauzac, priv. comm.) are marked on Figure~\ref{fig:optical-xray}, which shows an HST ACS/F814W image of the cluster, with overlaid \chandra\ and \emph{JVLA} contours. The centers of the X-ray cores were defined as the peaks in the X-ray emission, and were determined from the $0.5-4$~keV \emph{Chandra} surface brightness map of the cluster, convolved with a Gaussian of $\sigma=4\arcsec$. We also considered the uncertainties on the surface brightness distribution, and defined the uncertainties on the X-ray peaks as the regions around the cores in which the X-ray brightness is consistent (within the propagated Poisson errors on the counts image) with the brightness of the corresponding peak. We note that our approach only provides a lower limit on the extent of the peak regions, because the background was included in the surface brightness map used for this analysis. The uncertainties on the DM centers are $<1\arcsec$, significantly smaller than the uncertainties on the centers of the X-ray cores. In Figure~\ref{fig:DM-zoom}, we compare the position of the DM centers with that of the X-ray peak regions. Rather than confirming the previously-reported offset between the gas and DM components of the SW subcluster, Figure~\ref{fig:DM-zoom} shows that \emph{both} X-ray peaks are, within the uncertainties, co-located with the DM centers. We speculate that the discrepancy between the results presented herein and those reported by \citet{Jauzac2015} is caused by a higher noise level in the significantly shallower X-ray data used in previous studies; our data is $\approx 6$ times deeper, which is equivalent to a reduction in noise by a factor of $\approx 2.5$.

\section{Merger Scenario}
\label{sec:scenario}

\citet{Jauzac2015} showed that the galaxy redshift distribution in MACS~J0416.1-2403 is bimodal relative to the mean redshift, with the SW subcluster (mean redshift $0.3966$) moving towards the observer and the NE subcluster (mean redshift $0.3990$) moving away from the observer. Based on these observations, \citet{Jauzac2015} suggested two possible merger scenarios:
\begin{itemize}
        \item MACS~J0416.1-2403 is a \emph{pre-merging} system, in which the SW subcluster comes from behind the NE subcluster, and is now seen near first core passage;
        \item MACS~J0416.1-2403 is a \emph{post-merging} system, in which the SW cluster approached the NE subcluster from above, and is now seen near its second core passage.
\end{itemize}
In this section, we try to distinguish between the two scenarios based on our X-ray and radio findings. 

\subsection{Cooler gas in the NE?}

Approximately 300~kpc NE of the NE core, the temperature maps show a region (region \#10; $T=8.41_{-0.90}^{+1.37}$~keV) that is cooler than the neighboring region closer to the cluster center (region \#8; $T=12.18_{-1.90}^{+2.41}$). The cool gas in this region could be pre-shock gas ahead of a shock front. We examined the surface brightness profile across the cooler region, but could not confirm a surface brightness discontinuity with a confidence $\ge 90\%$. A density jump could be masked by projection effects. Nonetheless, while the temperature difference between regions \#8 and \#10 is significant at $90\%$ confidence (but not at $2\sigma$) and there is a pseudo-pressure jump between the two regions, the lack of a clear density jump does not allow us to confirm a shock front NE of the NE core.

\subsection{No gas-DM offset}

In head-on binary mergers, the essentially collisionless DM travels ahead of the collisional gas. There are several scenarios that could explain the spatial overlap of the DM and gas in MACS~J0416.1-2403:
\begin{enumerate}[(i)]
   \item  MACS~J0416.1-2403 is not a head-on merger that progresses outside the plane of the sky, and the NE and SW subclusters are seen before first core passage and have yet to interact strongly with each other.
   \item The NE and SW subclusters are seen after first core passage, and the DM and gas overlap only in projection. This requires either that one of the subclusters was essentially undisturbed by the merger and our line of sight aligns with the trajectory of the other subcluster, or that the subclusters' trajectories are parallel and both are aligned with our line of sight.
   \item The NE and SW subclusters are seen after first core passage, and the gas of the SW subcluster slingshot and caught up with the DM halo, after they had separated near pericenter.
\end{enumerate}  

The radial velocity difference between the BCGs of the NE and SW subclusters was reported by \citet{Jauzac2015} to be $\approx 800$~km~s$^{-1}$. The sound speed in a cluster with a temperature of $\approx 10$~keV is $\approx 1300$~km~s$^{-1}$, and typical collision velocities are $1-2$ times the sound speed. Therefore, the radial velocity difference between the two BCGs in MACS~J0416.1-2403 seems rather low. The velocity could be explained if the subclusters are seen well before or after first core passage, possibly near the turnaround point.


\subsection{Substructure in the NE core}

The surface brightness profile of the NE subcluster cannot be described by a single $\beta$-model, but instead by a double $\beta$-model composed of a very compact core and a more extended gas halo. Such a profile is typically used to model cool core clusters \citep[e.g.][]{JonesForman1984,Ota2013}, but it can also describe the surface brightness of merging clusters seen in projection \citep[e.g.][]{ZuHone2009,Machacek2010}. The core of the NE subcluster is very hot and has a long cooling time, which disfavors the possibility that the NE subcluster hosts a cool core. Instead, the observed substructure is more likely the result of a merger event.

The most straight-forward scenario would be that the NE and SW subclusters have already merged, and while the core of the SW subcluster was destroyed in the collision, as evidenced by the subcluster's flat surface brightness, the core of the NE subcluster survived. Cool cores that survive a recent merger event preserve part of their cold gas. However, we found no evidence of cool gas in the NE core. Therefore, the NE core is unlikley to be the remnant of a recent interaction between the NE and SW subclusters.

An alternative scenario is that the NE subcluster is merging with the dark matter halo S2 \ref{fig:unsharp}. However, S2 is not associated with a concentration of galaxies, and thus is an unlikely candidate for a group of galaxies; instead, S2 is probably a cosmic filament seen in projection along the line of sight \citep{Jauzac2015}. No other mass concentration that the NE subcluster might be currently merging with has been detected.

One other possibility, postulated by numerical simulations of \citet{Poole2008}, is that the NE subcluster underwent a merger or a series of mergers over $\sim 1$~Gyr ago, and while its core has already relaxed back to a compact state, it has not yet recooled. In this scenario, the merger could not have been with the SW subcluster, because the two subclusters are involved in an ongoing merger. Instead, the NE subcluster needs to have interacted several Gyr ago with one or more clusters that are no longer directly detectable in the X-ray maps, in the lensing maps, or in the redshift distribution of the NE subcluster. 

Alternatively, past mergers could also have heated the cool core without destroying it, by inducing sloshing in the central galaxy. The kinetic energy of the sloshing central galaxy would then have been dissipated as heat between successive mergers. 

\subsection{The AGN and the X-ray cavity in the NE core}

The collision velocity between the NE and SW subcluster is unlikely to be highly supersonic. If the cluster is a post-merging system, a cavity could have had enough time to form since first core passage. However, given the short timescale and the intense turbulence following the moment of first core passage, forming a new cavity would require a strong AGN outburst. The expectation would then be to observe radio emission associated with the X-ray cavity. However, this radio emission is not observed. Instead, the presence of an X-ray cavity NW of the NE core supports a pre-merger scenario. The cavity was most likely inflated by a recent weak outburst of the AGN detected in the BCG of the NE subcluster. 

\subsection{SW density discontinuity}

The density discontinuity detected S of the SW subcluster is located along the line connecting the NE subcluster, the SW subcluster, and S1. The discontinuity also appears coincident with the position of S1, but this could be a projection effect. There are two scenarios that would explain the origin of the density discontinuity: the merger of the SW subcluster with S1, or the merger of the SW subcluster with the NE subcluster.

If the discontinuity were triggered by a merger between the two main subclusters, then the merger must have already progressed past the moment of the first core passage. Furthermore, because the NE core is very compact and hosts an X-ray cavity, it is unlikely that it was significantly affected by a merger with the SW subcluster. Therefore, taking into account the constraints set by the lack of an offset between the gas and DM peaks of both subclusters, it is the SW subcluster whose trajectory must be aligned with our line of sight. In this geometry, a shock front would be expected ahead of the SW subcluster (with respect to its direction of motion), i.e. our line of sight should be perpendicular to the 3D shock ``cap''. Instead, however, the surface brightness discontinuity we detect is seen in projection only S-SW of the SW subcluster. If the detected density discontinuity is the appearance of the shock in projection, then we would expect to observe it over a larger angular opening. Most likely, the density discontinuity is caused by the interaction of the SW subcluster with S1.

\textbf{In conclusion, while we cannot completely eliminate the possibility of MACS~J0416.1-2403 being a post-merging cluster, we believe \emph{the sum of our findings favors a pre-merging scenario}.}

\section{A Newly-Discovered Radio Halo}
\label{sec:underluminous}

\begin{figure*}
	\includegraphics[width=\columnwidth]{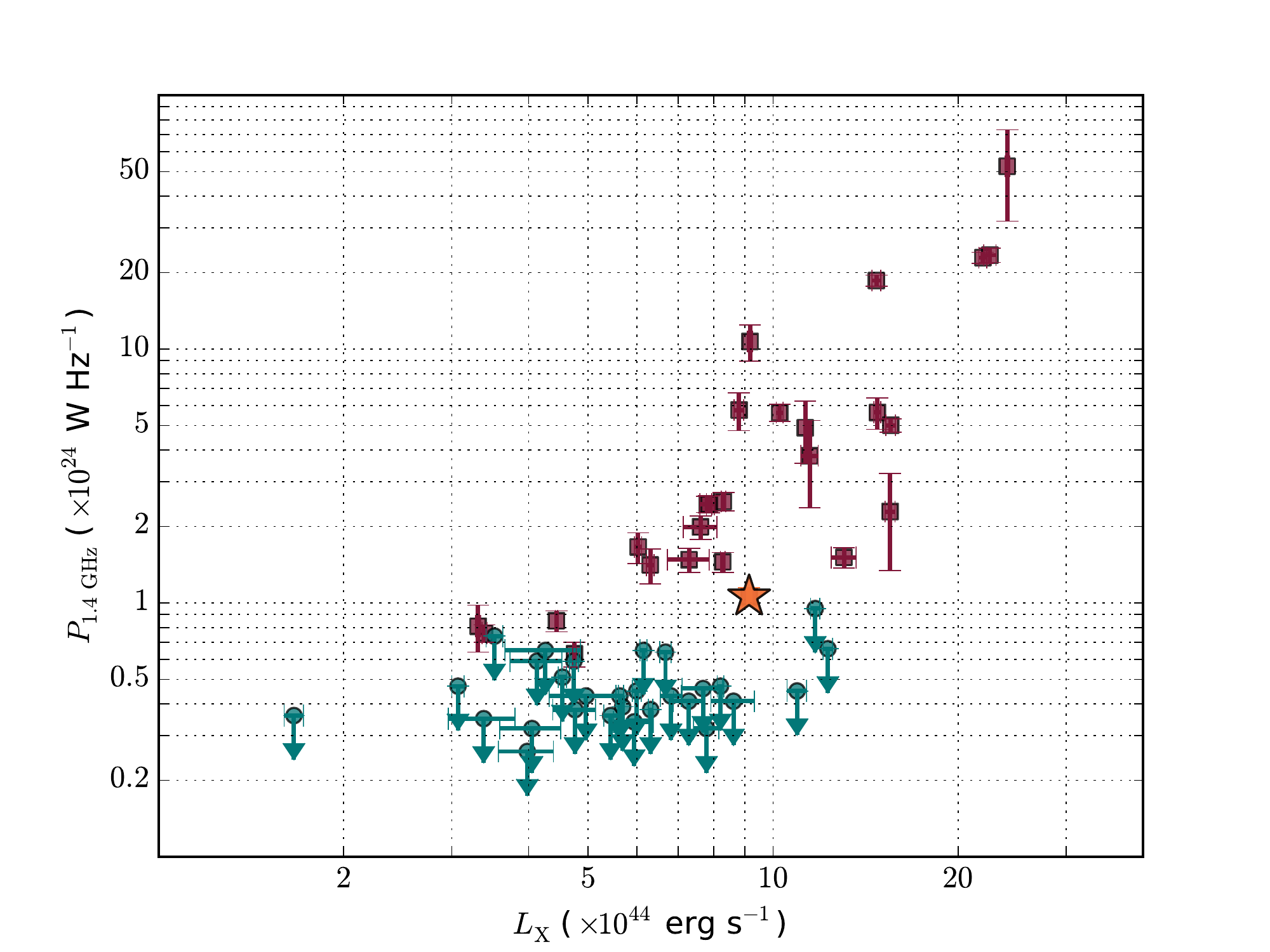}\hfill
	\includegraphics[width=\columnwidth]{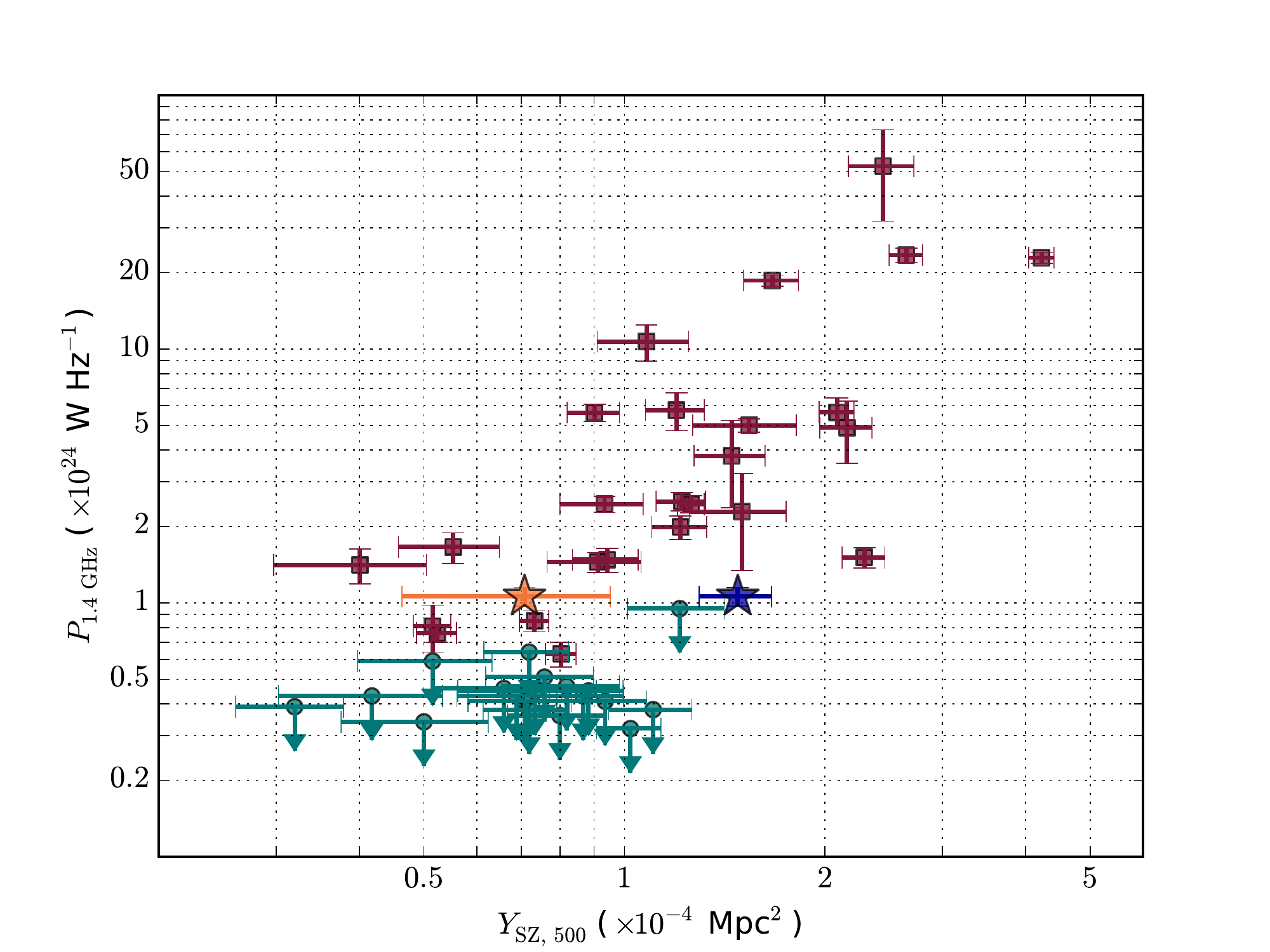}
	\caption{$L_{\rm{X}}-P$ and $Y_{\rm{SZ}}-P$ diagrams. Clusters with only upper limits on the radio halo power are shown in teal-colored circles, while clusters with detected halos are shown in dark red squares. On both diagrams, MACS~J0416.1-2403 is represented with a star. In the $Y_{\rm{SZ}}-P$ diagram, the orange star corresponds to the $Y_{\rm{SZ,\,\,500}}$ value reported by \citet{2015arXiv150201598P}, while the blue star corresponds to the $Y_{\rm{SZ,\,\,500}}$ value calculated from the gNFW fit to Bolocam presented in \citet{2014arXiv1406.2800C}. Error bars are drawn for all points, but some are very small and barely visible. Values for all clusters with the exception of MACS~J0416.1-2403 are taken from \cite{2013ApJ...777..141C}. An interactive version of the diagrams is available at \url{http://hea-www.cfa.harvard.edu/~gogrean/Lx-P.html}.\label{fig:underluminous}}
\end{figure*}

We classify the diffuse radio emission in the cluster as a halo, since the emission has low-surface brightness, large extent, and is centrally located. The halo has a somewhat smaller size than other radio halos in merging clusters, which have typical sizes of $1-1.5$ Mpc. \citep[e.g.,][]{Feretti2012}. The size is more similar to that of radio mini-halos found in cool-core clusters \citep[e.g.,][]{Giacintucci2014}. However, an important difference from mini-halos is that neither the NE nor the SW subclusters contains a cool core. In addition, as shown in Figure~\ref{fig:optical-xray}, the halo seems to be associated with both subclusters.

The radio emission around the northern subcluster contains about a factor of $2$ more flux than the emission around the southern one. An interesting question is whether we observe a single radio halo, caused by the merger between the NE and the SW subclusters, or two individual halos belonging to the two subclusters. In the latter case, these halos must have originated from  previous merger events within two subclusters themselves. This would make MACS~J0416.1-2403 similar to the case of the radio halo pair in Abell~399 and Abell~401 \citep{Murgia2010}, with the difference that Abell~399/401 seems to be at a much earlier pre-merger stage.

For radio halos, a correlation is observed between the cluster's X-ray luminosity and the radio halo power (the $L_{\rm{X}}-P$ correlation), with the most X-ray luminous clusters corresponding to the most powerful radio halos \citep[e.g.,][]{2000ApJ...544..686L,2002A&A...396...83E,2006MNRAS.369.1577C}.  The X-ray luminosity is used as a proxy of cluster mass.  The distribution in the $L_{\rm{X}}-P$ plane is bimodal when clusters without halos are included. Merging clusters with radio halos follow the $L_{\rm{X}}-P$ correlation, while the upper limits on the radio power of other clusters are well below the correlation \citep{2007ApJ...670L...5B}.

A correlation is also observed between the cluster's integrated Sunyaev-Zel'dovich (SZ) effect (i.e., the integrated Compton $Y_{\rm{SZ}}$ parameter) and the radio halo power \citep[e.g.,][]{2012MNRAS.421L.112B,2013AN....334..350B}. $Y_{\rm{SZ}}$ is proportional to the volume integral of the pressure.  The SZ signal should be less affected by the cluster dynamical state and therefore is expected to be a better indicator of the cluster's mass. \citet{2012MNRAS.421L.112B,2013AN....334..350B} and \citet{2013ApJ...777..141C} showed that there is good agreement between the $Y_{\rm{SZ}}-P$ scaling relation and the scaling relations based on X-ray data.

In Figure~\ref{fig:underluminous}, we place MACS~J0416.1-2403 on the $L_{\rm{X}}-P$ and $Y_{\rm{SZ}}-P$ diagrams, with the values for the other clusters taken from \cite{2013ApJ...777..141C}. The $Y_{\rm{SZ,\,\, 500}}$ value for the cluster was recently calculated by \citet{2015arXiv150201598P}, which found $Y_{\rm{SZ,\,\, 500}} = (0.71  \pm  0.24) \times 10^{-4}$~Mpc$^{2}$. Specifically, this is the value of $Y_{\rm{SZ,\,\, 500}}$ obtained from the union catalog based on the MMF1\footnote{Matched multi-filter algorithm, implementation 1.} detection algorithm. We also computed the value of $Y_{\rm SZ,\,\,500}$ using the gNFW model fit to Bolocam presented by \citet{2014arXiv1406.2800C} and obtained $Y_{\rm SZ,\,\,500} = (1.49\pm 0.28)\times 10^{-4}$~Mpc$^2$. 
The difference between the Planck and Bolocam measurements is not understood, but is likely due to some combination of random noise fluctuations, deviations from the assumed pressure profile used to extract $Y_{\rm{SZ,\,\, 500}}$ (particularly at large radii), contamination from dust or other astronomical signals, and calibration uncertainties. Furthermore, we note that this cluster is not detected by PowellSnakes -- one of the three Planck algorithms -- and the values of $Y_{\rm{SZ,\,\, 500}}$ obtained from the other two algorithms (MMF1 and MMF3\footnote{Matched multi-filter algorithm, implementation 3.}) are significantly different, perhaps indicating that measurements of $Y_{\rm{SZ,\,\, 500}}$ towards this cluster contain larger than typical systematic uncertainties. In Figure~\ref{fig:underluminous}, we plot both the Planck and Bolocam values of $Y_{\rm SZ,\,\, 500}$.

Interestingly, the radio power falls on the low end of the $L_{\rm{X}}-P$ correlation, which means that the radio halo in MACS~J0416.1-2403 is somewhat less luminous than expected from the X-ray luminosity of the cluster.
However, using the Planck measurement of $Y_{\rm{SZ,\,\, 500}}$, we find that the halo power is consistent with the expected $Y_{\rm{SZ}}-P$ correlation, indicating the halo is not underluminous for the cluster mass. While the Bolocam value of $Y_{\rm{SZ,\,\, 500}}$ does imply lower than expected halo power, the nominal $Y_{\rm{SZ}}-P$ relation used for comparison is derived solely from Planck SZ measurements. We therefore consider our interpretation based on the Planck value of $Y_{\rm{SZ,\,\, 500}}$ to be more robust. Given this interpretation, MACS~J0416.1-2403 could be just an outlier in the $L_{\rm{X}}-P$ diagram, based on the fact that there is significant scatter in the correlations. 

It is also possible that the X-ray luminosity of MACS~J0416.1-2403 is not a good indicator of the cluster's mass, and the luminosity is boosted by the merger. A boost in X-ray luminosity is predicted for approximately one sound-crossing time, which for MACS~J0416.1-2403 is $\approx 1$ Gyr \citep{Ricker2001}. This boost would affect other clusters on the correlation as well, but it could affect MACS~J0416.1-2403 more since we are seeing a cluster close to first core passage.  The mass of the cluster in $R_{\rm 500}$ is $(7.0\pm 1.3)\times 10^{14}$~M$_{\sun}$ \citep{Umetsu2014}. Based on the scaling relations of \citet{Pratt2009}, the mass corresponds to a $0.1-2.4$~keV cluster luminosity of $(1.1\pm 0.4)\times 10^{45}$~erg~s$^{-1}$, which is in agreement with the luminosity calculated from the \emph{Chandra} data. Therefore, based on the $M_{\rm 500}-L_{\rm 500}$ scaling relation, we have no indication that the cluster luminosity is higher than expected for a cluster of this mass.

Another possibility is that we are seeing an ultra-steep spectrum radio 
halo \citep[$\alpha \lesssim -1.5$,][]{2008Natur.455..944B}. In this case, 
the radio halo is only underluminous at higher frequencies (i.e., above a GHz), 
while the integrated flux density rapidly increases at lower frequencies. The $320-360$~MHz VLITE 
and 610~MHz GMRT data presented in Section~\ref{sec:jvla-results} are 
not sufficient to confirm, or rule out, the presence of an ultra-steep spectrum radio halo. Low-frequency GMRT ($\nu < 610$~MHz) and/or 
deep P-band JVLA observations are required to settle this case.

A third possibility is that we could be witnessing the first stages of the formation of this radio halo. We do not think that the radio halo is fading, because the NE and SW subclusters are still in the process of merging and most likely they have not yet undergone core passage (see Section~\ref{sec:scenario}). Therefore, when the merger progresses further, the radio halo could brighten, increase in size, and move onto the correlation.

\section{Summary}
\label{sec:summary}

The HST Frontier Field cluster MACS~J0416.1-2403 ($z=0.396$) is a merging system that consists of two main subclusters and two additional smaller mass structures \citep{Jauzac2015}. Here, we presented results from deep \chandra\, JVLA, and GMRT observations of the cluster. The main aims of our analysis were to identify signatures of merger activity in the ICM, constrain the merger scenario, and detect and characterize diffuse radio sources. Below is a summary of our main findings:
\begin{itemize}
	\item The cluster has an average temperature of $10.06_{-0.49}^{+0.50}$~keV, an average metallicity of $0.24_{-0.04}^{+0.05}$~Z$_{\sun}$, and a $0.1-2.4$~keV luminosity of $L_{\rm X}=(9.14\pm 0.09)\times 10^{44}$~erg~s$^{-1}$.
	\item The NE subcluster has a very compact core and a nearby X-ray cavity, but its temperature is very high ($T\sim 10$~keV), we find no evidence of multi-phase gas, and its cooling time is $\gtrsim 3$~Gyr. The presence of a compact core and of an X-ray cavity indicates that the NE subcluster was not disrupted by a recent major merger event.
	\item S-SW of the SW subcluster, there is a weak density discontinuity that is best-fitted by a density compression of $1.56_{-0.29}^{+0.38}$. The discontinuity is located along the line connecting the NE subcluster, the SW subcluster, and the mass structure S1 reported by \citet{Jauzac2015}. Most likely, the discontinuity was caused by a collision between the SW subcluster and S1, rather than by a collision between the NE and SW subclusters.
	\item The DM and gas components of the NE and SW subclusters are well-aligned, which favors a scenario in which the two subclusters are seen before first core passage.
	\item MACS~J0416.1-2403 has a radio halo with a 1.4~GHz power of $(1.06\pm 0.09)\times 10^{24}$~W~Hz$^{-1}$, which is somewhat less luminous than predicted by the $L_{\rm X}-P$ correlation. However, the halo aligns well on the $Y_{\rm SZ,\,\,500}-P$ correlation, indicating that it is not underluminous for the cluster mass. We could be observing the cluster at a point in the merger event when the X-ray luminosity is significantly boosted. Alternatively, we could be observing the birth of a giant halo, or a very rare ultra-steep halo.
\end{itemize}



\acknowledgments

We thank Mathilde Jauzac for providing the coordinates of the centers of the DM halos, and 
Aurora Simionescu for constructive discussions about the X-ray analysis.

GAO acknowledges support by NASA through a Hubble Fellowship grant HST-HF2-51345.001-A awarded 
by the Space Telescope Science Institute, which is operated by the Association of Universities 
for Research in Astronomy, Incorporated, under NASA contract NAS5-26555. RJvW is supported 
by NASA through the Einstein Postdoctoral grant number PF2-130104 awarded by the Chandra X-ray Center, 
which is operated by the Smithsonian Astrophysical Observatory for NASA, under contract NAS8-03060. 
AZ acknowledges support by NASA through a Hubble Fellowship grant HST-HF2-51334.001-A awarded by the 
Space Telescope Science Institute, which is operated by the Association of Universities for 
Research in Astronomy, Incorporated, under NASA contract NAS5-26555. This research was performed 
while TM held a National Research Council Research Associateship Award at the Naval Research 
Laboratory (NRL).  Basic research in radio astronomy at NRL by TM and TEC is supported by 6.1 Base funding.

The scientific results reported in this article are based on observations made by 
the \chandra\ X-ray Observatory. Part of the reported results are based on observations 
made with the NASA/ESA Hubble Space Telescope, obtained from the Data Archive at the Space 
Telescope Science Institute, which is operated by the Association of Universities for Research 
in Astronomy, Inc., under NASA contract NAS 5-26555. The HST observations are associated with 
program \#13496. The National Radio Astronomy Observatory is a facility of the National 
Science Foundation operated under cooperative agreement by Associated Universities, Inc.
We thank the staff of the GMRT that made these observations possible. The GMRT is run by 
the National Centre for Radio Astrophysics of the Tata Institute of Fundamental Research.

This research has made use of NASA's Astrophysics Data System, and of the NASA/IPAC 
Extragalactic Database (NED) which is operated by the Jet Propulsion Laboratory, 
California Institute of Technology, under contract with the National Aeronautics and 
Space Administration. Some of the cosmological parameters in this paper were calculated 
using Ned Wright's cosmology calculator \citep{2006PASP..118.1711W}.



{\it Facilities:} \facility{Chandra (ACIS)}, \facility{HST (ACS)}.




\bibliography{macsj0416-chandra}
\clearpage




\end{document}